*Irradiation of TiBe12 and CrBe12 with He+ ions and ex-situ transmission electron microscopy to investigate ion damage as a proxy for neutron irradiation damage in nuclear fusion*

Jo Sharp[1]*, Viacheslav Kuksenko[2], Ramil Gaisin[3], Graeme Greaves[1], Jonathan Hinks[1], Pavel Vladimirov[3], Stephen Donnelly[1]

1: School of Computing and Engineering, University of Huddersfield, Queensgate, Huddersfield, HD1 3DH, UK

2: United Kingdom Atomic Energy Authority, Culham Science Centre, Abingdon, Oxfordshire OX14 3DB, United Kingdom

3: Karlsruhe Institute of Technology, Hermann-von-Helmholtz-Platz 1, 76344 Eggenstein-Leopoldshafen, Germany

# 1. Abstract

Titanium and chromium beryllides, $TiBe_{12}$ and $CrBe_{12}$, are materials of potential future importance as neutron multipliers for tritium breeding in nuclear fusion reactors. Beryllium experiences extremely high transmutation according to a n →2n transmutation reaction in which both tritium and helium are produced, which normally form bubbles in solids at the relevant concentration range. Neutron irradiation from the fusion plasma also introduces point defects into solids. The ensuing effect of this environment on the beryllides' microstructure is poorly characterised, but important for understanding beryllides' mechanical properties and their evolution in the irradiative environment inside a fusion reactor. This study is intended to initially determine and describe the microstructural features that occur in $TiBe_{12}$ and $CrBe_{12}$ when He and fast-particle-induced point defects have been introduced at fusion reactor neutron breeder relevant temperatures. In this study, beryllide samples were implanted with 300kV He at a range of temperatures between 387-900°C, sectioned down through the implantation surface with a focused ion beam post-irradiation, and examined the resulting microstructures using transmission electron microscopy, electron-dispersive spectroscopy (EDS) and precession diffraction mapping.

Nanometre-scale bubbles grew in both $TiBe_{12}$ and $CrBe_{12}$ at 600°C and larger (100+ nm) bubbles, some faceted, grew at 900°C. Some bubbles in $CrBe_{12}$ were lined with Cr, with some of the Cr oxidised. $TiBe_{12}$ developed planar faults, on {110} planes at 600°C and below but on to {111} at 900°C. Faults were preferentially associated with large bubbles. The displacement vectors of faults on the {110} planes had some commonality with previous studies that found displacement vectors of the two types $\boldsymbol{R} = \frac{1}{2}\langle 011 \rangle$ and $\boldsymbol{R} = \frac{1}{2}\langle 110 \rangle$; the present study also found faults that did not match either previously found type. $CrBe_{12}$ also developed planar faults but the appearance of these was quite different from the typical striped appearance of planar stacking faults and their nature remains unknown. Oxide particles were found in both beryllides, most prominently in $CrBe_{12}$.

# 2. Introduction

Beryllium stands to become an important element in the construction of nuclear fusion reactors, including as a part in test blanket modules to trial designs for tritium breeding structures. Be's role is as a neutron multiplier due to a set of nuclear reactions between high-speed neutrons and Be nuclei that produce more neutrons, $^6$He, $^4$He and tritium $^3$H [1] [2]. The main problem for Be as a neutron multiplier material, however, also arises from these reactions: the He isotopes produced form bubbles, which grow at the high temperatures expected in the breeder module, cause swelling of the part, and trap He and tritium in the bubbles or in their walls, [1], [2], [3]. This also structurally weakens the Be and results in Be dust, which makes an extra hazard to dispose of when parts are changed. Tritium is radioactive and a large inventory of it in bubbles inside parts complicates maintenance and disposal of damaged parts. Using Be pebbles to give a large surface area and small internal volume alleviates this a little but the pebbles still show some gas retention, swelling and corresponding weakening of mechanical properties [4].

High-energy neutron irradiation trials of beryllide intermetallics such as $TiBe_{12}$, $CrBe_{12}$ have shown lower tritium and helium retention and less swelling at temperatures up to 600°C, with higher temperature performance hitherto unknown [5]. A breeder made of beryllide material should therefore improve performance in terms of gas retention and structural integrity. The structure of a number of $MBe_{12}$ beryllide intermetallics including Ti and Cr is ordered body-centred tetragonal, with the transition metal M at the lattice points (0,0,0) and unit cell centre, and Be atoms in the following three non-equivalent sites: Be1 (0.25, 0.25, 0.25), Be2 (0.361, 0, 0); Be3 (0.277, 0.5, 0). The lattice parameters of $TiBe_{12}$ ($CrBe_{12}$ in brackets) are *a*=*b*=7.3328Å (7.19395Å), *c*=4.145Å (4.123Å) and the space group is $I\frac{4}{m}mm$ [6]. Some density functional theory simulation studies found that the most energetically favourable vacancy is at the Be2 position, and that two neighbouring Be2 vacancies (a divacancy) can capture up to 12 He, or 5 He + 4 H [6], though this is not yet a settled question.

A second phase, $M_2Be_{17}$, is also found in some systems, such as Ti-Be, but not others, such as Cr-Be. Its structure is monoclinic, with cell parameter c being close but not necessarily the same as a and b, and the non-orthogonal angle nominally 120°. The $M_2Be_{17}$ structure is made by the addition of an extra (001) plane into the $MBe_{12}$ structure, a shift of $\frac{1}{2} < 110 >$, and some small shifts of Be sites to minimise stress. The crystallographic axes used to describe $M_2Be_{17}$ are defined in a different orientation to those of $MBe_{12}$; [001] of $MBe_{12}$ is in the (001) plane of $M_2Be_{17}$ perpendicular to [010] [7]. The structure of $M_2Be_{17}$ thus has somewhat lower symmetry. Figure 1 illustrates the two structures: a1 and a2 are views of $MBe_{12}$ viewed down [100] and [001] respectively; b is $M_2Be_{17}$ viewed down the [001] axis.

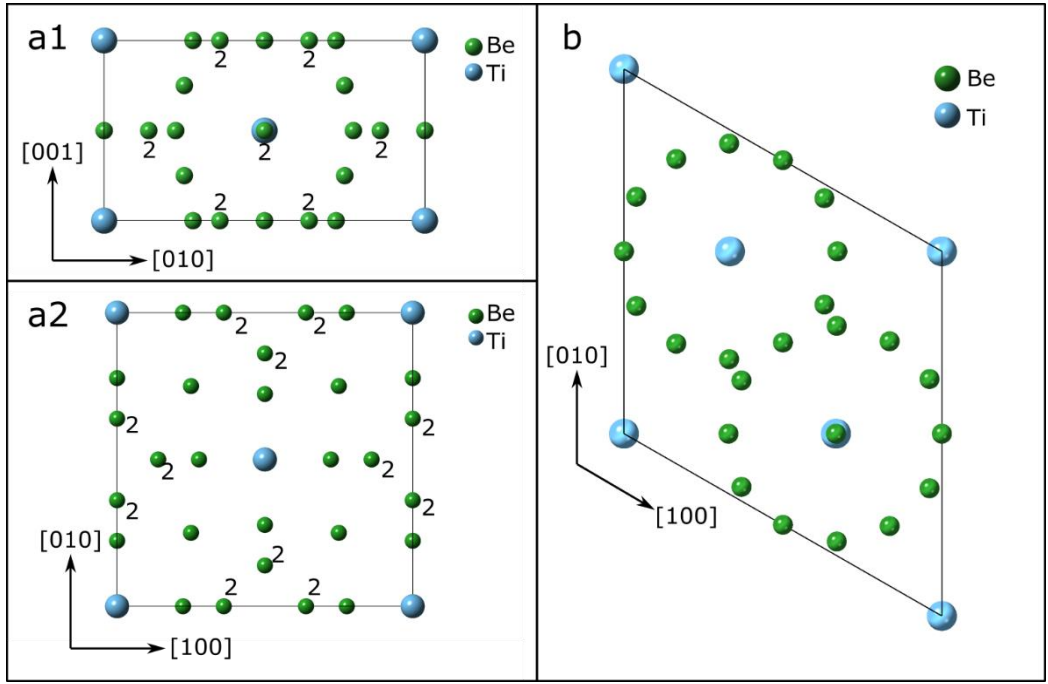

*Figure 1 Structures of MBe$_{12}$ and M$_2$Be$_{17}$ structures, where green is Be and blue is M. Image (a1): MBe$_{12}$ from [100] direction with Be2 type sites labelled. (a2): MBe$_{12}$ structure from [001] with Be2 sites labelled. (b): M$_2$Be$_{17}$ structure from [001].*

Experimental studies of the microstructure of MBe$_{12}$ are few and far between; Banerjee et. al. found a number of structural defects:

1. A superlattice incorporating stripes of second phase Ti$_2$Be$_{17}$ in which the TiBe$_{12}$ lattice is disturbed every sixth (001) plane (a spacing of 6 x 4.145 Å ≈ 2.5 nm) by displacement vector $\frac{1}{2}[011]$ creating a single layer of Ti$_2$Be$_{17}$ unit cells. This produces satellite diffraction spots attached to the $g = 011$ type spots, displaced in the [002]* reciprocal lattice direction.
2. Dislocations of Burgers vector $\boldsymbol{b} = \frac{1}{2}<111>$ with image contrast described as complex.
3. Antiphase boundaries, curved in shape, with displacement vector $\boldsymbol{R} = \frac{1}{2}<110>$.
4. Twinning on {101} planes.

Beryllides have been produced only since the latter half of the 20$^{th}$ century [8], a short history compared to better studied materials. Various production methods have been used which affect the microstructure of the end product. In the study above, arc melting produced the TiBe$_{12}$/Ti$_2$Be$_{17}$ superlattice mentioned, and areas of pure TiBe$_{12}$ with dislocations and antiphase boundaries; hot isostatic pressing (HIP) of atomised material produced deformation twins [7]. In more recent work comparing the density and phase purity of the resulting material between arc-melting and powder-HIP methods, HIP at 1100°C was found better on both counts and chosen for future beryllide manufacture for fusion applications [9].

The purpose of this research is to find an important piece of information not yet known about the MBe$_{12}$ intermetallics: the microstructural features and events that occur under irradiation at various temperatures. This knowledge is key to understanding why they retain less He and tritium and how to optimise their performance and use, and this work is intended to begin filling that knowledge gap. The co-authors at Karlsruhe Institute of Technology worked on the long term HIDOBE project to irradiate Be and beryllide pellets to high neutron doses over 6 years, in a more realistic simulation of the fusion reactor environment; post-irradiation examination of those radioactive samples also requires complex, costly and time-consuming procedures. In this work we have carried out low-cost, short, high-flux He ion irradiation as an analogue for high-cost, long-duration neutron irradiations such as HIDOBE. We present the results here in order to enable assessment of the similarities and differences between the two experiment types, and the caveats that must be placed on the results of He implantation experiments when using them to assess material suitability for the fusion environment.

## 3. Method

The samples of $TiBe_{12}$ and $CrBe_{12}$ for this study were provided by the Karlsruhe Institute of Technology, prepared using vacuum hot pressing of beryllide powder. The grain size of $TiBe_{12}$ before irradiation had a bimodal distribution, with sizes in the ranges 5-10 μm and 20-40 μm [10]. The original grain size of $CrBe_{12}$ was ~40μm.

Slices of $TiBe_{12}$ and $CrBe_{12}$ of approximate dimensions 20 x 20 x 2μm were lifted out by focused ion beam (FIB) at the Materials Research Facility at UKAEA, attached to Mo support grids. The slices were then irradiated using the MIAMI-2 system at the University of Huddersfield [11], by 300keV He ions at a flux of 1.0-1.2 x $10^{13}$ ions/cm$^2$/s for durations of 1h (fluence ~4.2 x $10^{16}$ ions/cm$^2$) at temperatures 387°C, 480°C, 600°C and 900°C in a Gatan™ double tilt heating holder– the three lower temperatures correspond to time-average measured temperatures of the three sample locations in the HIDOBE-2 study [12]; 900°C was chosen to probe the effects on beryllides of irradiation well above the HIDOBE temperature range. After irradiation, perpendicular sections were lifted from the irradiated slices (onto Cu grids for cost reasons), such that each section contained a cross-section normal to the original incident surface along through the He travel direction, including the depth of peak He implantation. This is referred to as the "re-FIB" method in this paper. Conventional transmission electron microscopy (TEM) was then performed using a Hitachi H-9500 LaB$_6$ source TEM at 300kV. Further analysis was carried out using scanning TEM (STEM), including electron-dispersive X-ray spectroscopy (EDS/EDX) mapping 200kV on the JEOL F200 TEM at the University of Sheffield.

The thickness of 2μm and He energy of 300kV were chosen to place the predicted peak ion implantation depth approximately halfway through the beryllide slab in order to lessen surface effects, modelled using the *Stopping and Range of Ions in Matter* (SRIM) Monte Carlo simulation code [13]. The "quick" method in SRIM was used according to Stoller's recommendation [14]; the displacement energy $E_d$ of these materials is not yet known, so SRIM calculations were carried out using $E_d$ of 20eV and 100eV, assuming the unknown accurate value is somewhere within that range, and both profiles were contained suitably within 2μm thickness. Resulting plots of damage in displacements per atom (dpa), ppm He and the ratio between them are given in Figure 2. The appm He curve is identical for both range-end $E_d$ values.

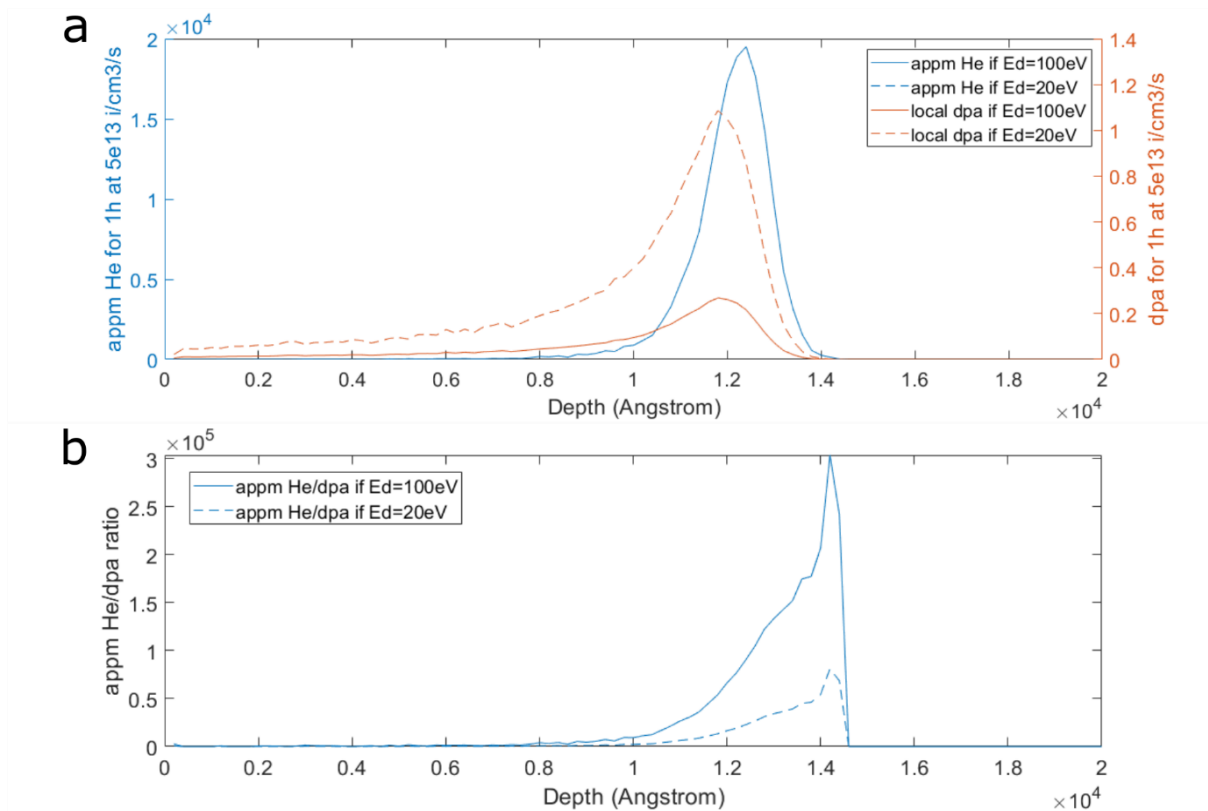

*Figure 2 Damage plots calculated using SRIM. (a) appm He (left axis) and dpa (right axis) for approximately the time and flux used; the appm He plot is the same for both displacement energies. (b) Resulting ratios of (appm He)/(dpa)*

When determining whether bubbles are present during transmission electron microscopy (TEM), the usual method is to take an image at focus and then change the focus to both under and over; if bubbles are present, their edges will give Fresnel fringes, with the nearest fringe to the bubble edge being bright in underfocus conditions (Δf -ve) and dark in overfocus conditions (Δf +ve) e.g. [15].

The $TiBe_{12}$ sample irradiated at 387°C was characterised by precession diffraction mapping using the NanoMEGAS ASTAR system, indexing to Be, $MBe_{12}$ and $M_2Be_{17}$ structures but not accounting for strain at this stage. Crystal structures .cif files were obtained or compiled from the following sources: Be from the library structure in CrystalMaker from [16]; $TiBe_{12}$ from the structure given in [6]; $Ti_2Be_{17}$ from dataset mp-1201649 on The Materials Project [17]. Precession data in the NanoMEGAS sytem has a corresponding "reliability" parameter, a measure of the confidence attached to the decision to assign a given point as belonging to the assigned phase; it is the inverse of the relative strength of the next-most-likely candidate structure during phase fitting to the diffraction pattern taken from that point. This has been superimposed onto the precession phase maps; a brighter point is more confidently indexed as its assigned phase, usually because it is near a zone axis; a darker point is more ambiguous between the candidate phases, perhaps because it is far from a zone axis with few points in the diffraction pattern. Strain also makes indexing less reliable, moving diffraction spots away from their model positions.

## 4. Results

In this section, the terms "upstream" and "downstream" refer to the direction of ion travel in the original thick specimen during ion irradiation; downstream relative to an object means away from it along the direction of ion travel, upstream relative to an object means back towards the irradiated surface.

## 4.1 Ti beryllide samples

### 4.1.1. Sample RLB 1: TiBe$_{12}$ irradiated with 300keV He at 387°C

#### *4.1.1.1 Summary*

This sample shows speckling in TEM images on the lower half of the cross section and close to the irradiated surface, which is found by EDS and precession diffraction mapping to be not an irradiation-related feature but a FIB artefact of redeposited Be and Cu from the support grid. No bubbles are seen.

A planar fault on a $(1\bar{1}0)$ habit plane is found; long curved dislocations are also found nearby close to the peak He concentration depth.

#### *4.1.1.2 Speckling; absence of bubbles*

A low magnification montage is shown in Figure 3a; the material further into the cross section than the He concentration peak is speckled, while the irradiated material above the He concentration peak is smooth. There is also some speckling near the entrance surface where less He is expected to stop. A through-focus set of images from the He concentration peak depth is shown in Figure 3b; a range of defocus values and magnifications was used to look for Fresnel fringes that would indicate bubbles, but none are seen – either there are none or they are sub-nm and too small to see in the available magnification range.

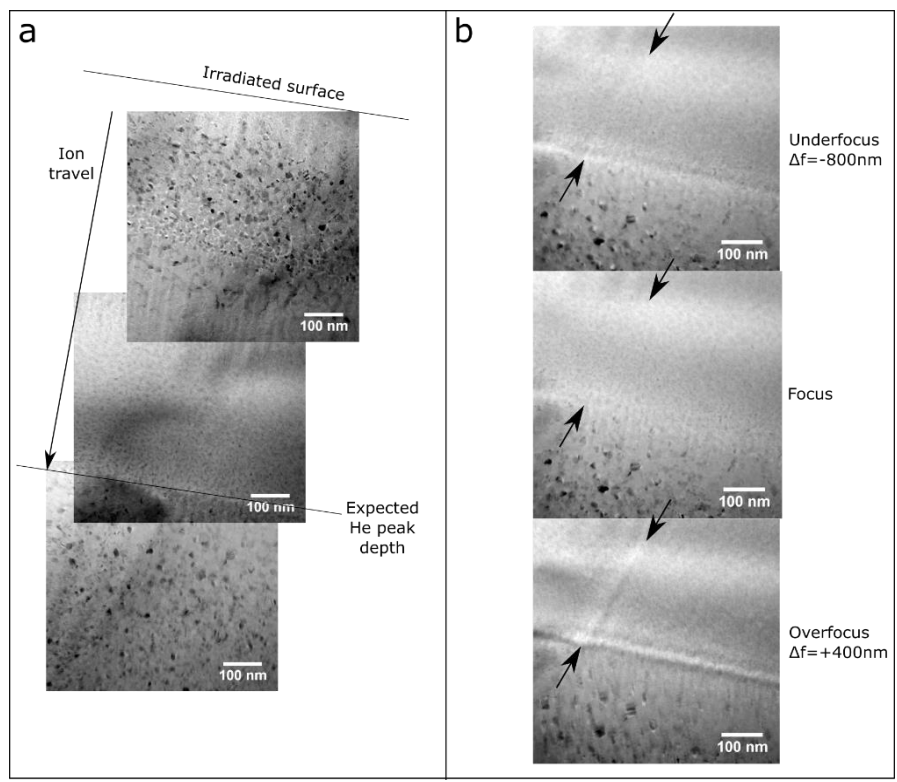

*Figure 3 (a) Montage of bright field images through TiBe$_{12}$ sample irradiated at 387°C. Irradiated surface and depth of peak implanted He concentration (calculated using SRIM) are marked. (b) A selection of the images either side of focus used to identify bubbles. A planar fault is indicated by arrows.*

High-angle annular dark field scanning TEM (HAADF-STEM) images are shown in Figure 4. Image (a) shows the speckled region below the irradiated surface, with a region expanded in image (c). Image (b) shows at higher magnification the unirradiated speckled region below the He concentration peak. The only element whose map corresponds to the speckles in the HAADF STEM image is a slight correlation in the Cu map, shown in Figure 4d. This implies adherence of Cu resputtered from the grid during FIB.

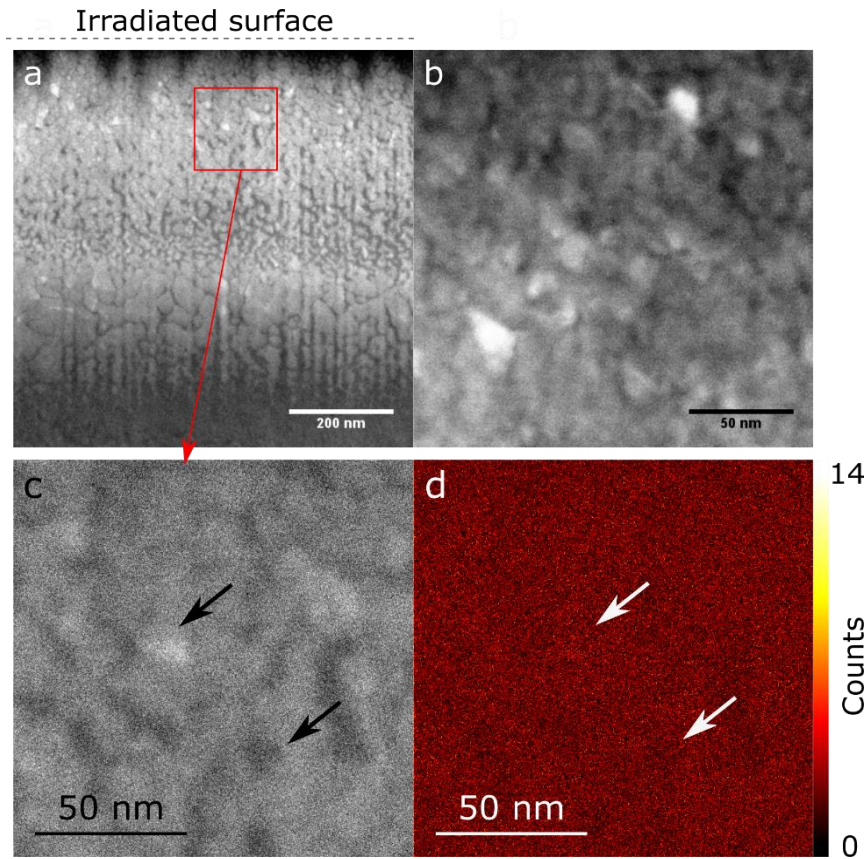

Figure 4 HAADF-STEM images from TiBe$_{12}$ sample irradiated at 387°C. (a) Speckled contrast under the irradiated surface – annotation shows the approximate position of the irradiated surface. (b) small-grained unirradiated region beneath the He zone. (c) higher magnification HAADF-STEM image of the region in the red box in (a). (d) Cu EDS map of the region in (c) showing a slight correlation of Cu with the speckling, for example the light and dark patches marked by arrows.

In order to find out whether the material under the redeposition is single-phase TiBe$_{12}$ and whether it was single-grained or polycrystalline, precession diffraction mapping was employed, giving the results in Figure 5. Selected area diffraction patterns match the previously expected structure of TiBe$_{12}$, body centred tetragonal. Traditional selected area diffraction using a selected area aperture identifies this grain to be near the [001] zone axis, as do the nanodiffraction patterns from the non-speckled part of the sample such as that in Figure 5a; during original He irradiation, the He ion beam would have been near to [$\bar{1}\bar{1}0$]. The speckled unirradiated region is indexed as Be by the ASTAR software; diffraction patterns from this region such as that shown in Figure 5d, however, contain the same TiBe$_{12}$ [001] grain diffraction pattern as in the smooth areas, with additional diffraction spots of stronger intensity from the redeposited phase (arrowed red in 5d and 5c). This indicates that the TiBe$_{12}$ under the redeposition is a single grain. This overlapping pattern is only a few spots, so the indexing has a lower certainty; it may have been mis-indexed as Be and the redeposition may actually be TiBe$_{12}$, and the Cu identified by EDS. This is also the case for the speckled region at the top, shown in diffraction pattern Figure 5c. Diffraction pattern (b) is referred to in the next subsection.

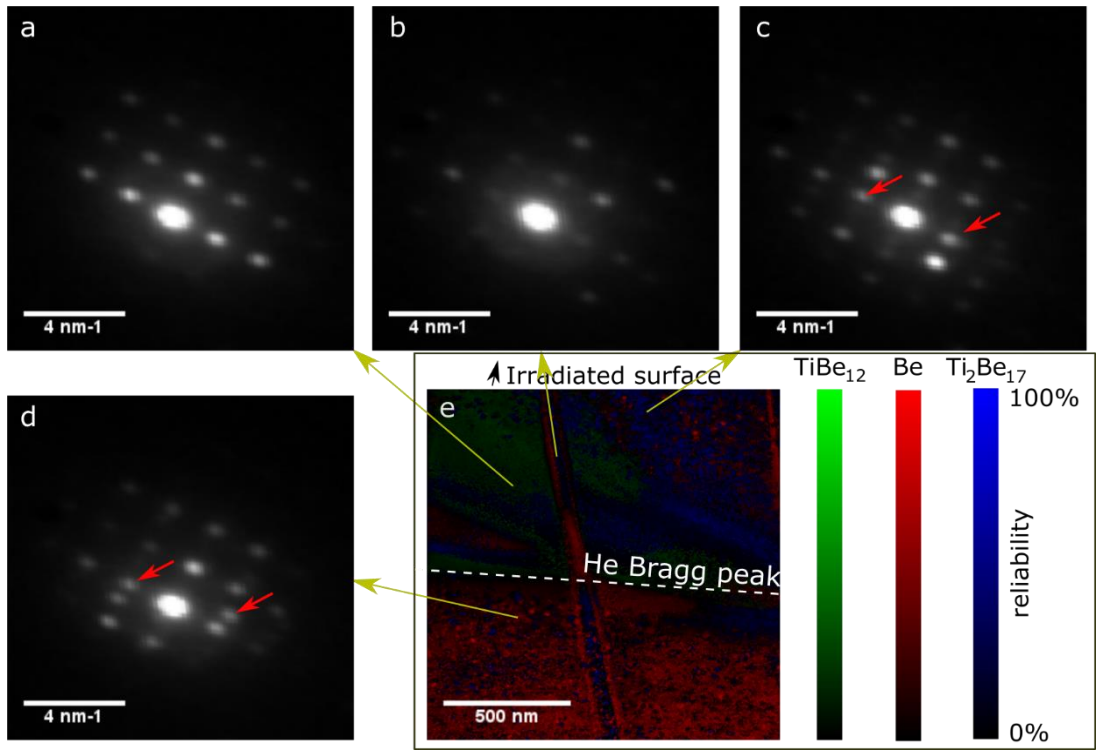

*Figure 5 Precession diffraction phase mapping from around a fault in the TiBe$_{12}$ sample irradiated at 387°C. (a) Pattern from non-speckled area indexed as TiBe$_{12}$; base [001] pattern. (b) Pattern from planar fault, indexed as Ti$_2$Be$_{17}$; this is the base TiBe$_{12}$ pattern with some spots missing. (c) Pattern from speckled area near irradiated surface, TiBe$_{12}$ [001] base pattern with Be pattern superimposed. (d) Pattern from speckled area in unirradiated material downstream from the He concentration peak, also the TiBe$_{12}$ [001] base pattern combined with Be pattern. (e) Precession phase map indexed by Nanomegas software. The rotation of the diffraction patterns with respect to the image is not calibrated in this figure.*

### 4.1.1.3 Planar fault and dislocations

Conventional TEM dark field images of the planar fault found in Figure 3b are shown in Figure 6. The fault lies close to the $(1\bar{1}0)$ habit plane; this is not the same habit plane as the regularly spaced faults reported by Banerjee et. al., which were on the (001) plane [7] with that habit plane critical to the structural relation between TiBe$_{12}$ and Ti$_2$Be$_{17}$. The planar fault shown in Figure 6 is therefore not of the variety seen in [7]. The diffraction patterns either side of the fault in Figure 6 are identically aligned, so this fault is also not a grain boundary, but some other kind of planar defect. Long curved dislocations can be seen associated with the fault near the He concentration peak depth, which are strongly visible in the $g = 1\bar{1}0$ and g = 200 dark field images of Figure 6.

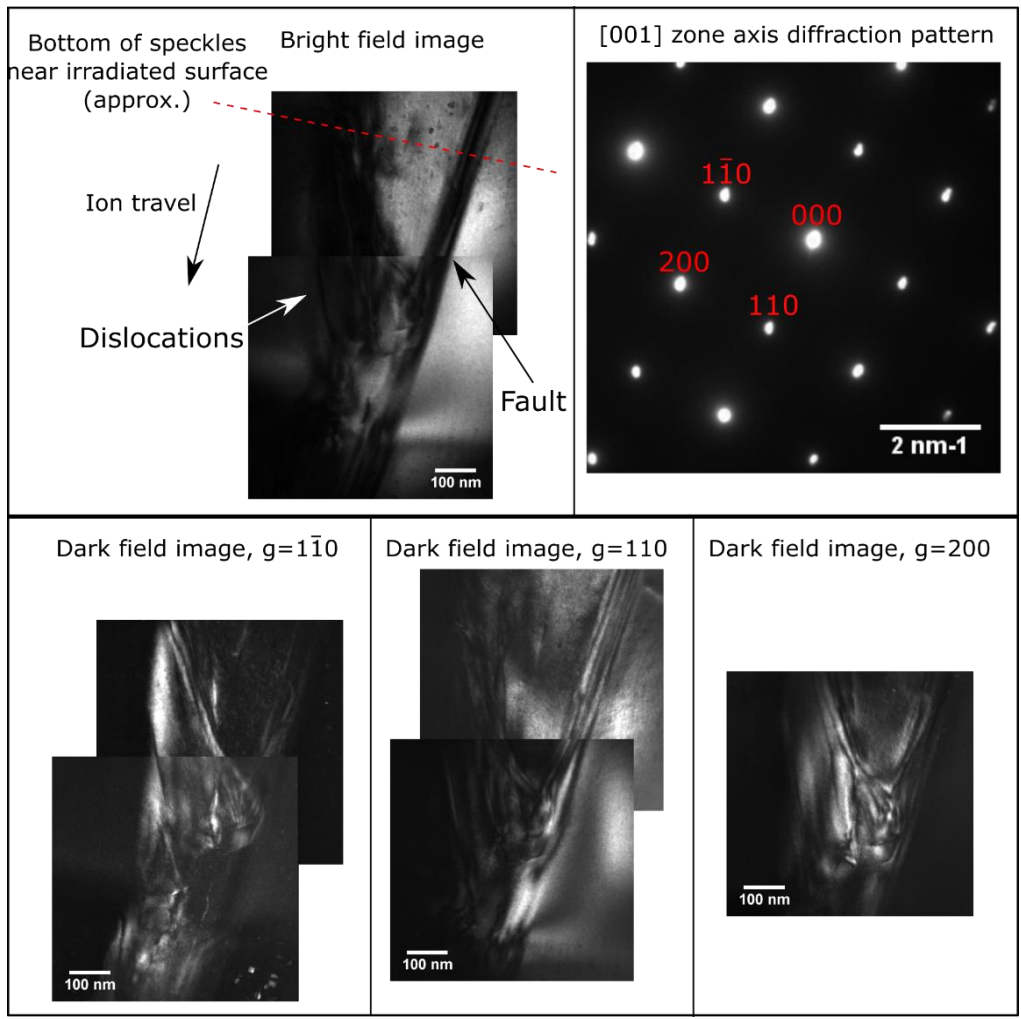

*Figure 6 A planar fault in the 387°C irradiated TiBe$_{12}$ sample taken using BF and different DF spots. The diffraction pattern is rotationally aligned with respect to the images. The orientation of the sample in these images is the same as Figure 3 with the original irradiated surface towards the top of the page.*

This planar fault is also shown in the precession diffraction map of Figure 5. There appears to be a band of Ti$_2$Be$_{17}$ structure along the planar fault running vertically down the centre of the precession phase map in Figure 5e (coloured blue), similar to the planar defects found by Banerjee et. al. [7] despite being on a different habit plane. A diffraction pattern from the fault is shown in Figure 5b, which has the same spot positions as the base TiBe$_{12}$ [001] pattern but with different patterns of spot intensity. This could be a mis-index due to the intensity distribution, or this could genuinely be Ti$_2$Be$_{17}$ with a fully coherent interface (i.e. plane spacings in common) in an orientation relationship with the surrounding TiBe$_{12}$. There also appears to be an area of Ti$_2$Be$_{17}$ in the top right of precession phase map Figure 5e; this region is bent or tilted off zone and the diffraction pattern (not shown) only has a few spots, so this region is probably bent TiBe$_{12}$ that has been mis-indexed due to a sparse diffraction pattern.

It is important to note the following limitations when interpreting this precession diffraction dataset: 1) strain is not accounted for in this initial dataset, and the arrangements of bend contours around the He zone indicate that this is a very strained sample, so precession work on irradiated beryllides requires more complex processing than this preliminary exploration; 2) this is carried out on one small piece of a little-studied material and repeat and expanded characterisation is required in future.

### 4.1.2. Sample RLB 2: TiBe$_{12}$ irradiated with 300keV He at 480°C

#### 4.1.2.1 Summary

This sample shows less redeposition than the 387°C sample. Bubbles are not seen at the He peak depth. Multiple parallel planar faults are found, also on the $(1\bar{1}0)$ habit plane.

#### 4.1.2.2 Planar faults

A bright field low magnification montage is shown in Figure 7. The sample is on the [001] zone axis (Figure 7b) and the montage shows an isolated fault at one side of the sample (A) and multiple parallel faults at the other side (B). Images at a range of defocus values and magnifications were taken at the area where the isolated fault A intersects the He concentration peak; no bubbles are seen, though some contrast is observed from surface Cu contamination (identified by EDS mapping). The set of parallel faults B crosses a grain boundary between the larger grain that makes up most of the sample and a grain which is captured at one corner of the sample. These two grains have an orientation relationship such that zone axis [101] of the main grain is parallel to zone axis [001] of the corner grain (Figure 7c).

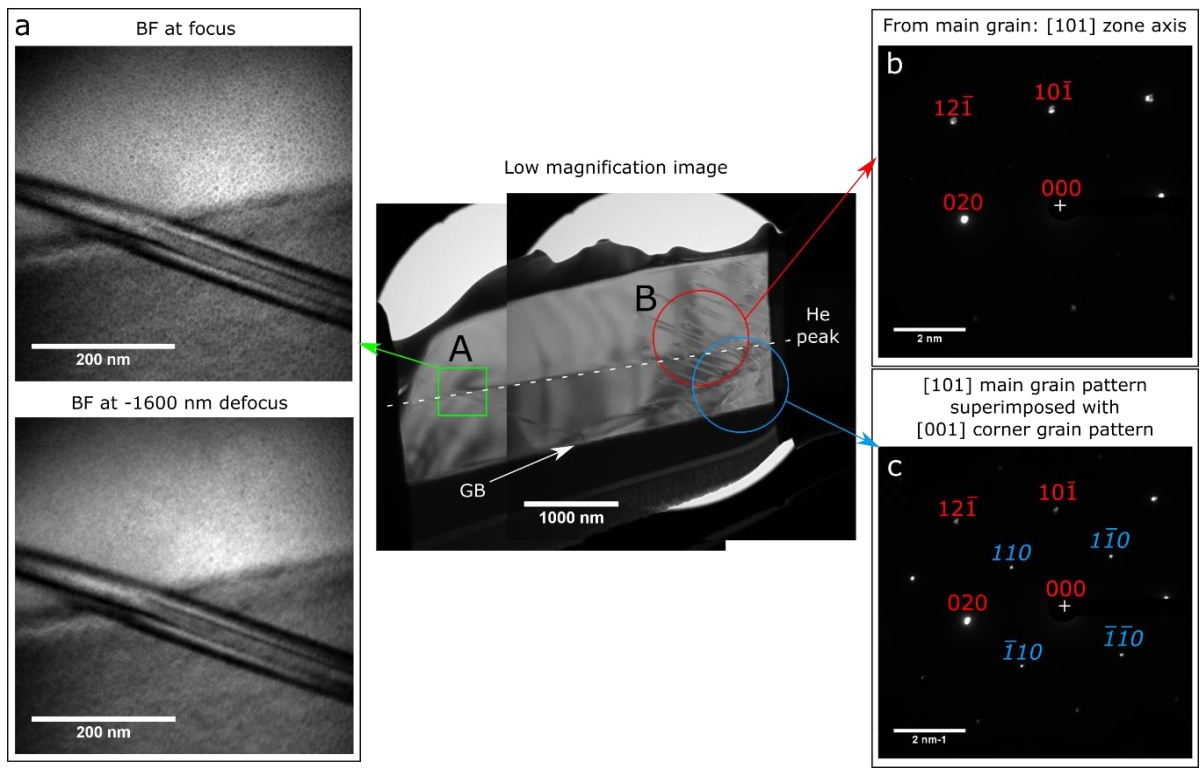

*Figure 7 Low magnification bright field image montage of the 480°C irradiated TiBe$_{12}$ section. The lone fault that crosses the He zone is enlarged on the left at focus and -1600nm defocus. The diffraction patterns on the right were taken using selected area apertures placed at the locations circled on the image; the main grain's spots are indexed in red and normal text, the corner grain's spots are indexed in blue and italic.*

The parallel faults are shown in bright and dark field in Figure 8, corresponding to the red circle at B in Figure 7. At the ends of the straight faults are bowed linear features, which diffract strongly to $g = 020$ (black lines) and $g = 12\bar{1}$ (white lines) and are barely visible in $g = 10\bar{1}$. The fault planes themselves can be determined from their orientation relative to the rotation-compensated diffraction pattern, and they align with the expected traces of $(1\bar{1}0)$ planes, the same as those in the 387°C sample.

Nanodiffraction was also performed on one of the faults in the [101] grain, well clear of the grain boundary, shown in Figure 8 c and d. The fault locally contains the zone axis [001] as well as [101], as if it is a continuation of the [001] corner grain that it protrudes from. The intersections of the faults with the grain boundary appear in Figure 8b, e, f and g to occur at steps in the grain boundary. Modelling of different fault types in this structure will take place in future work to find out what kind of defect would produce these observations.

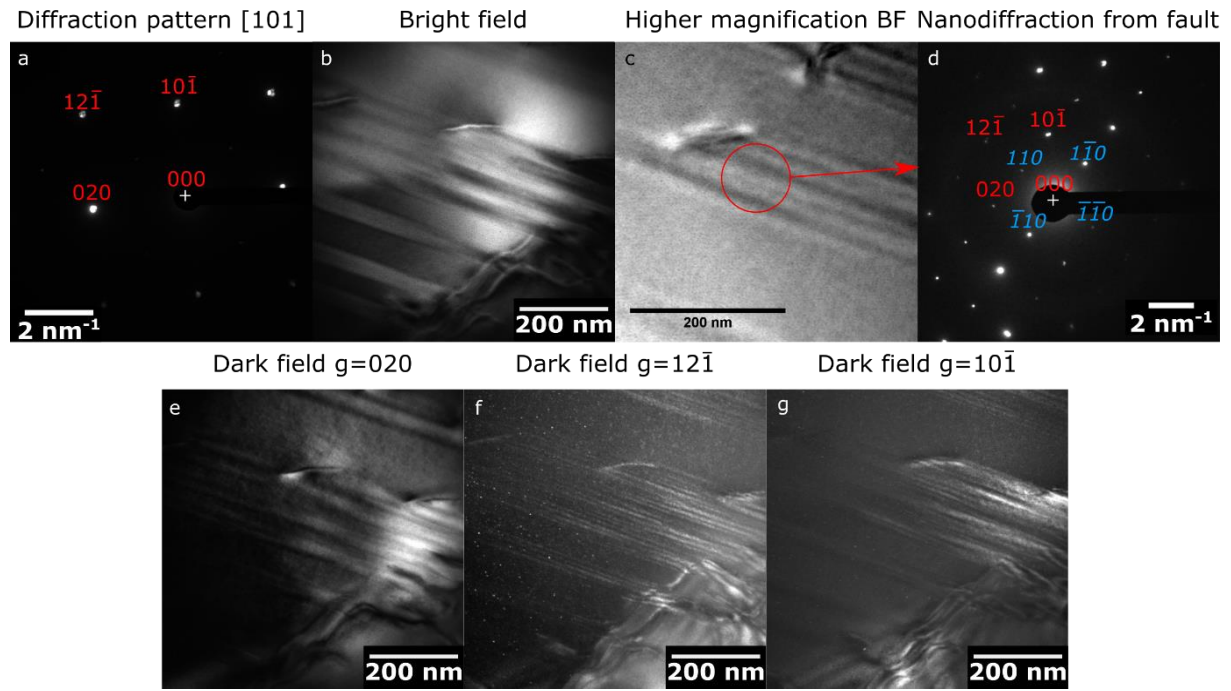

*Figure 8 Image set from heavily faulted area in 480°C irradiated TiBe$_{12}$ sample. (a) Diffraction pattern from main grain on [101] zone axis. (b) Bright field image of faults. (c) Higher magnification image of faults (actual location just out of image b to the left). (d) Nanodiffraction pattern from faults at point indicated in (c) – on the fault, the [001] zone axis is also present. (e-g) Dark field images of faults in g=020, $g = 12\bar{1}$ and $g = 10\bar{1}$ respectively.*

**4.1.3.** Sample RLB 3: TiBe$_{12}$ irradiated with 300keV He at 600°C

*4.1.3.1  Summary*

Nanoscale bubbles are found at the He concentration peak in this sample. A set of intersecting planar faults are found on non-parallel planes in the {110} family.

*4.1.3.2  Bubbles*

A bright field TEM montage of this sample is shown in Figure 9, showing a strong distortion in bend contours at the implanted He concentration peak compared to the lower temperatures, indicating more internal stress around the He concentration peak to cause bending when the re-FIB sample was cut. A band of ~1 nm scale bubbles are found at the He concentration peak by imaging either side of focus, also shown in Figure 9.

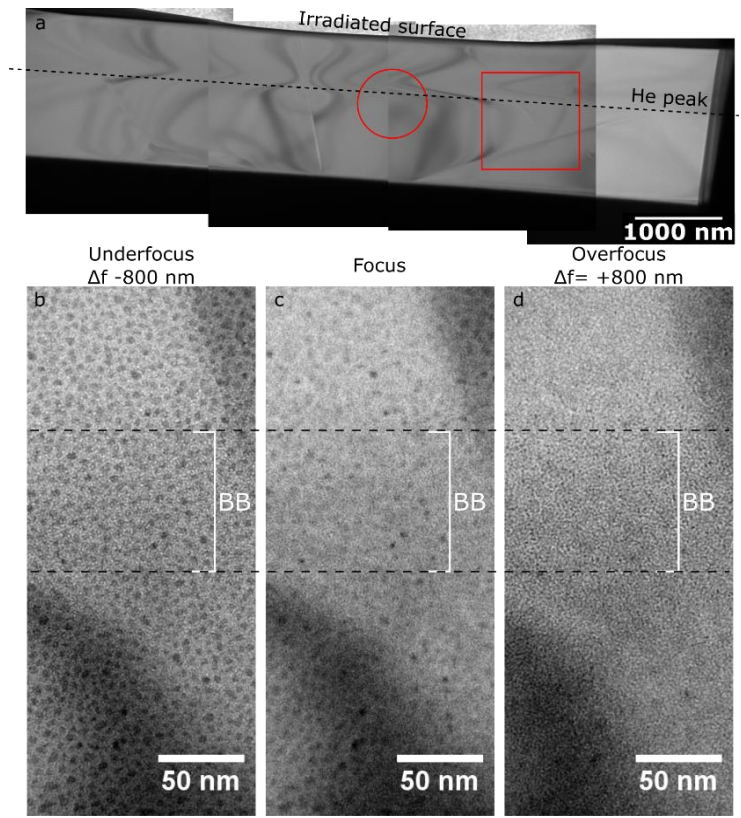

*Figure 9 (a) Low magnification bright field image of 600°C irradiated TiBe$_{12}$ specimen, not on zone. The red square shows the area where the fault complex in Figure 10 is found; the circle shows the origin of the diffraction pattern in that figure. (b)-(d) Images of the band where bubbles are most visible at the He concentration peak (annotated BB) at defocus values of -800nm (underfocus), focus, and +800nm (overfocus). The coarse dark speckle is contamination; the bubbles are the ~1nm speckles that invert contrast either side of focus indicated in the bracketed depth range.*

### 4.1.3.3  Planar faults

This sample too contains a number of planar faults, not parallel but an arrangement of multiple faults in the area highlighted by the red square in Figure 9a. Bright and dark field imaging was carried out on this arrangement with the sample on zone, which is the [315] zone axis, shown in Figure 10. From the angles of the faults relative to each other and the diffraction pattern, the fault planes are of type {110}, as for the TiBe$_{12}$ samples irradiated at 387°C and 480°C.

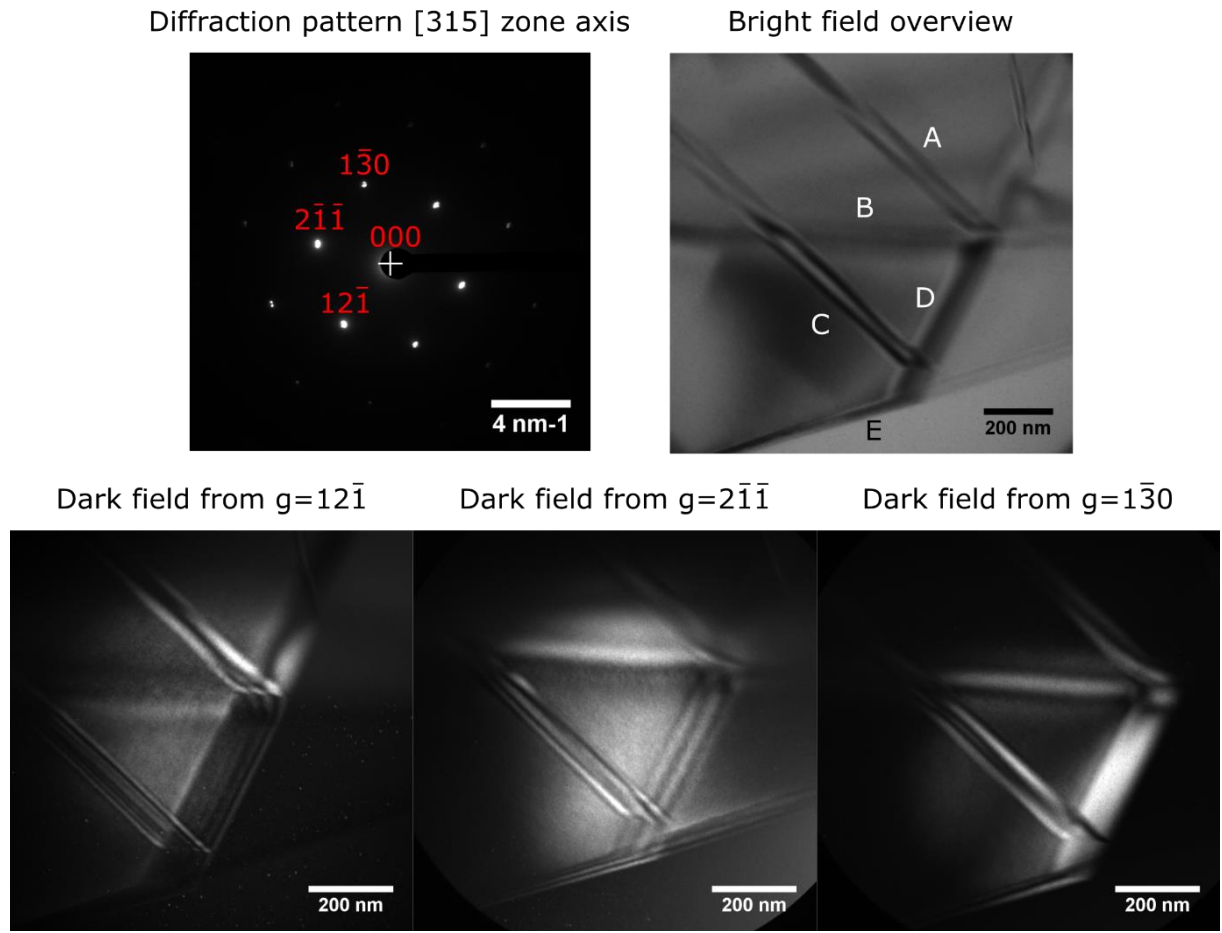

*Figure 10 Set of BF/DF images from a fault complex in the 600°C irradiated TiBe$_{12}$ sample. Letters A-E in bright field sample label individual faults in the complex for analysis in the Discussion section.*

**4.1.4.** Sample RLB 4: TiBe$_{12}$ irradiated with 300keV He at 900°C

*4.1.4.1 Summary*

This sample has major differences to those irradiated at the lower temperatures. Large (50-200nm) bubbles are found on grain boundaries and inside grains, with some nucleated on oxide particles. Grain boundary bubbles are generally larger than bubbles inside grains.

A high density of planar faults are found on {111} type planes, different from the {110} faults found before. These faults are preferentially associated with bubbles inside grains.

*4.1.4.2 Bubbles*

This sample, shown in the low magnification off-zone montage in Figure 11a, shows bubbles near grain boundaries (arrowed red) and in grain interiors. The He concentration peak is marked by the dashed line. The bubbles on the He concentration peak marked by box b in Figure 11a are explored off-zone in the focus set of Figure 11b. Bubble A, fairly round with diameter ~200 nm, is at a grain boundary triple point (marked with green dotted lines in the focus image). The other bubbles seen here are faceted in shape and offset by 50-200 nm from the nearest grain boundary. Striped contrast from faults can also be seen at many of the grain boundaries, such as at B, and in some of the matrix.

A set of bubbles or voids upstream of the He concentration peak, marked by box c in Figure 11a, are shown in another focal set in Figure 11c. It is not certain whether these are bubbles or voids; they are away from the He concentration peak, and it was not possible in this dataset to obtain a high enough energy resolution local electron energy loss (EELS) spectrum to determine definitively whether there is He inside. Here again there appear to be two populations: grain boundary bubbles/voids, spheroid in shape and ~200nm in diameter, such as those labelled C and D, and smaller geometric/faceted bubbles/voids 50-200 nm from the nearest grain boundary. The other bubbles/voids (right of image) are again geometric in shape and 50-200 nm from the nearest grain boundary. Here the grain boundary bubbles/voids C and D are accompanied by attached arrays of faults, shown by arrows in the +1600nm defocus image. This does not appear to be the case for grain boundary bubble A at the He concentration peak; this could indicate these upstream objects have more defect movement associated with them, but bubble A is also on a triple junction between grain boundaries which are themselves decorated with fault arrays, so the triple point position may be a contributor to the difference in fault association. The area at the right hand side of this sample downstream of the He concentration peak where bubbles/voids are also visible follows the same pattern with two types of bubble/void in association with faults.

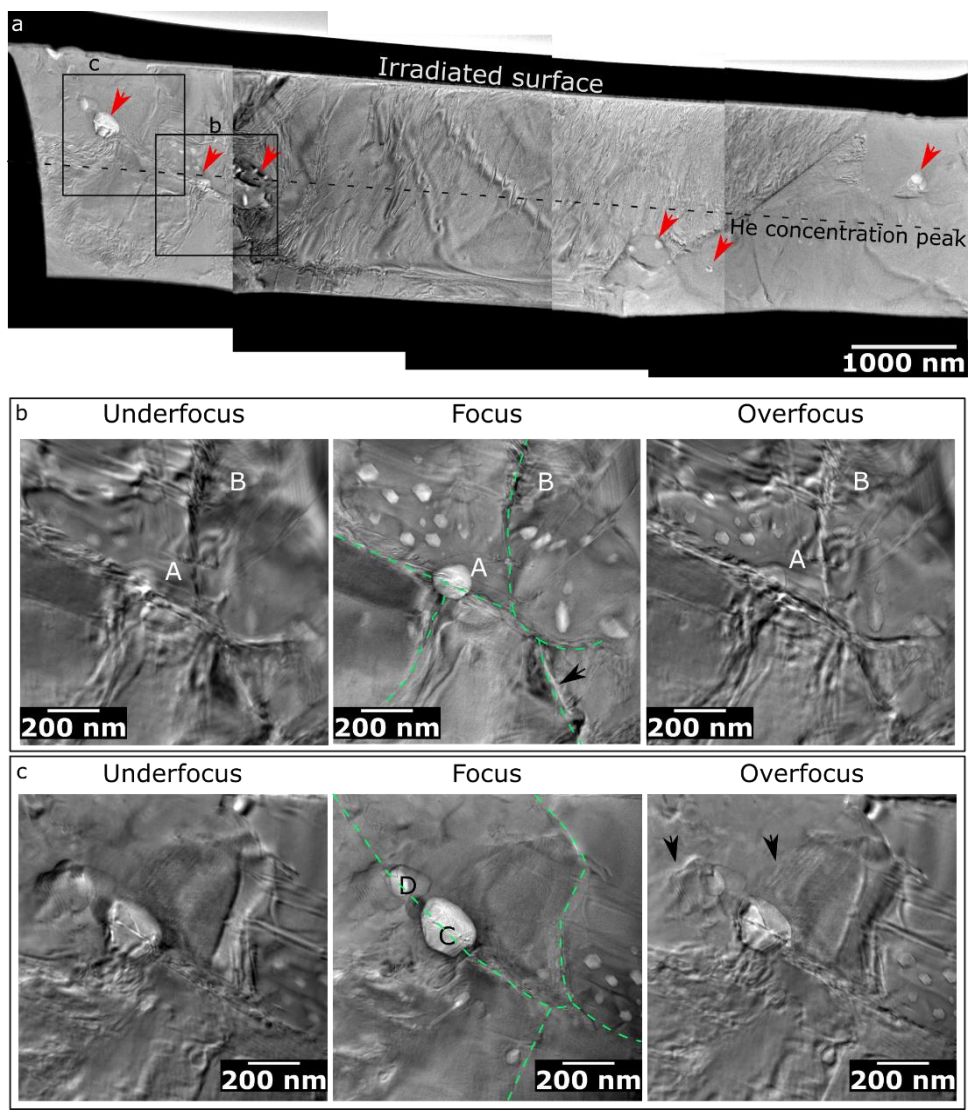

Figure 11 (a) Low magnification off-zone montage of 900°C irradiated TiBe$_{12}$ sample. Large bubbles are arrowed red. The dashed line marks the approximate position of the He concentration peak. (b) Bubbles on the He concentration peak. Grain boundaries are marked with green dashed lines in the focus image. Planar faults can be seen on the grain boundaries, for example at B. (c) Bubbles or voids upstream of He concentration peak. Grain boundaries are marked with green dashed lines. Overfocus and underfocus are +/-1600nm defocus.

Figure 12a shows STEM images and EDS maps from the region of box 1 in Figure 15, showing bubbles on the grain boundary in the unirradiated region downstream of the peak in implanted He concentration. Only the oxygen EDS map shows any contrast other than holes at the bubbles – the bright object is a grain boundary oxide precipitate on which a bubble has nucleated. The two grains that are brighter in ADF-STEM are also brighter in the Ti-K$_\alpha$ EDS map but this is probably orientation contrast. Bubbles on the opposite side of the TEM specimen in and above the bubble-rich region around the He band are shown in Figure 12b, which is from box 2 in Figure 15. No chemical segregation was found by EDS, including Ti, in these bubbles also.

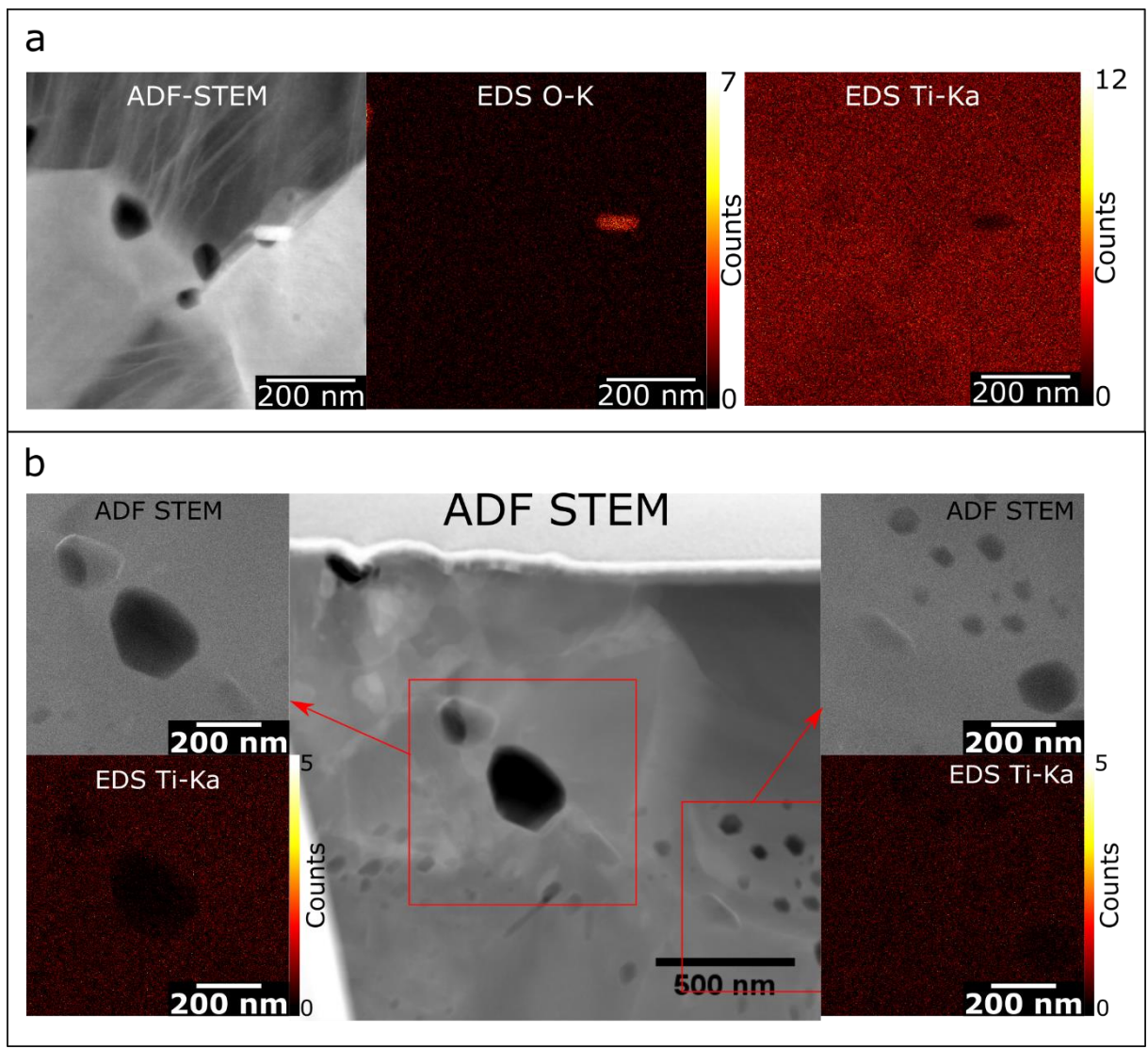

*Figure 12 (a) ADF-STEM and O-K and Ti-Kα EDS map from 900°C irradiated TiBe$_{12}$ sample, of grain boundary bubbles in unirradiated material below the He concentration peak. This region is marked 1 in Figure 15. (b) Bubbles in and upstream of the peak He concentration band. EDS maps show no Ti segregation around bubbles. This region is marked 2 in Figure 15.*

### 4.1.4.3 Planar faults

Faults in the interior of the central heavily faulted grain were imaged in bright field and dark field, shown in Figure 13. The arrangements of faults observed here follow a pattern of increasing complexity as irradiation temperature increases, though with the limitation that these specimens are small and not necessarily statistically representative of the whole. The zone axis here is [111]; the {110} type fault planes seen at lower temperatures would be either edge on or inclined at 22° to the plane of the sample and would intersect with each other at different angles seen projected onto the zone axis. Faults on {111} fit better the projected intersection angles seen in these images. This means the fault plane and possibly atomic arrangements at the faults have changed between 600°C and 900°C. Further analysis is ongoing to extract more information about the faults.

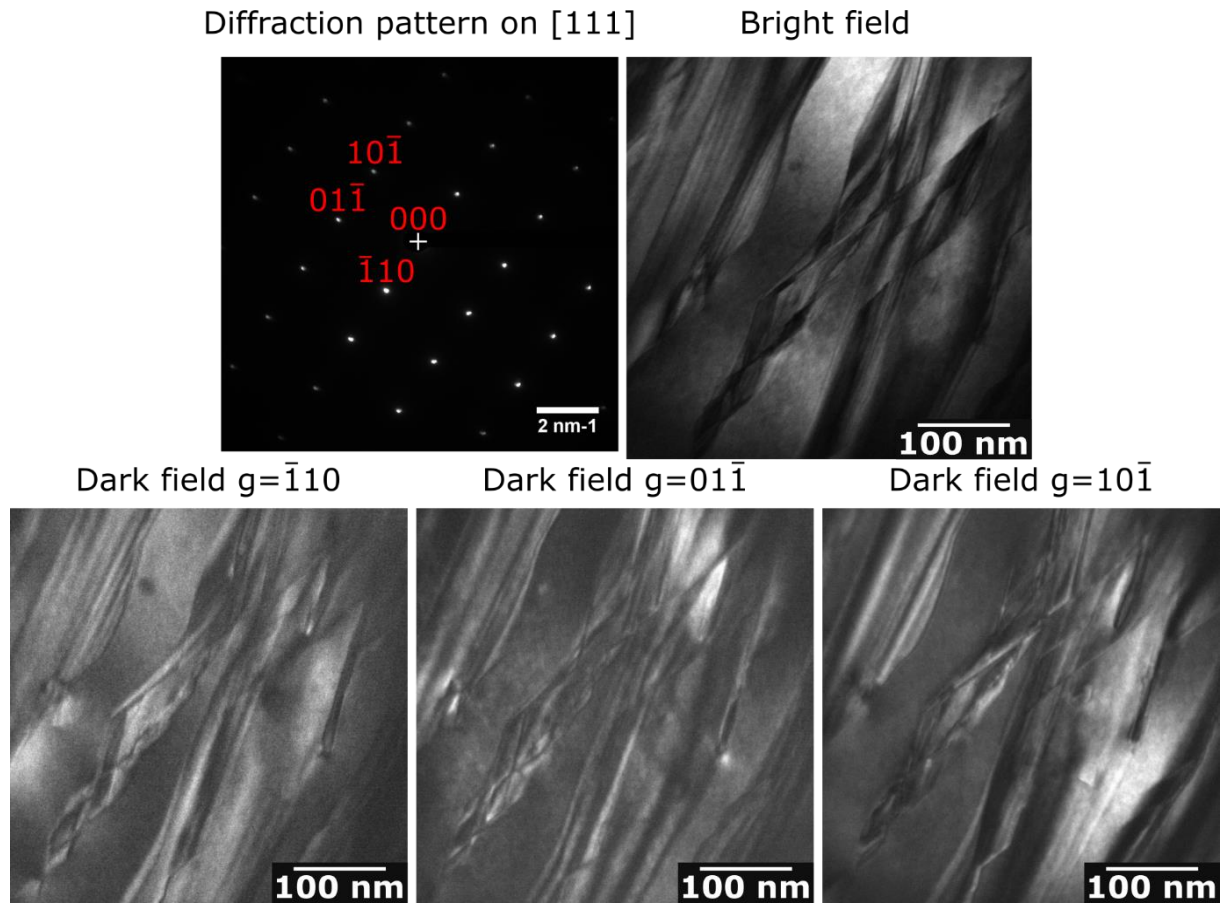

*Figure 13 BF/DF image set from central heavily faulted grain of 900°C irradiated TiBe$_{12}$ sample.*

In the central faulted grain is one set of parallel faults which produces superlattice spots between $g = 10\bar{1}$ type spots in every alternate row that follows that reciprocal lattice direction, shown in Figure 14. The indices of these spots are non-integer, e.g. $g = \frac{3}{2}, \bar{1}, \frac{\bar{1}}{2}$.

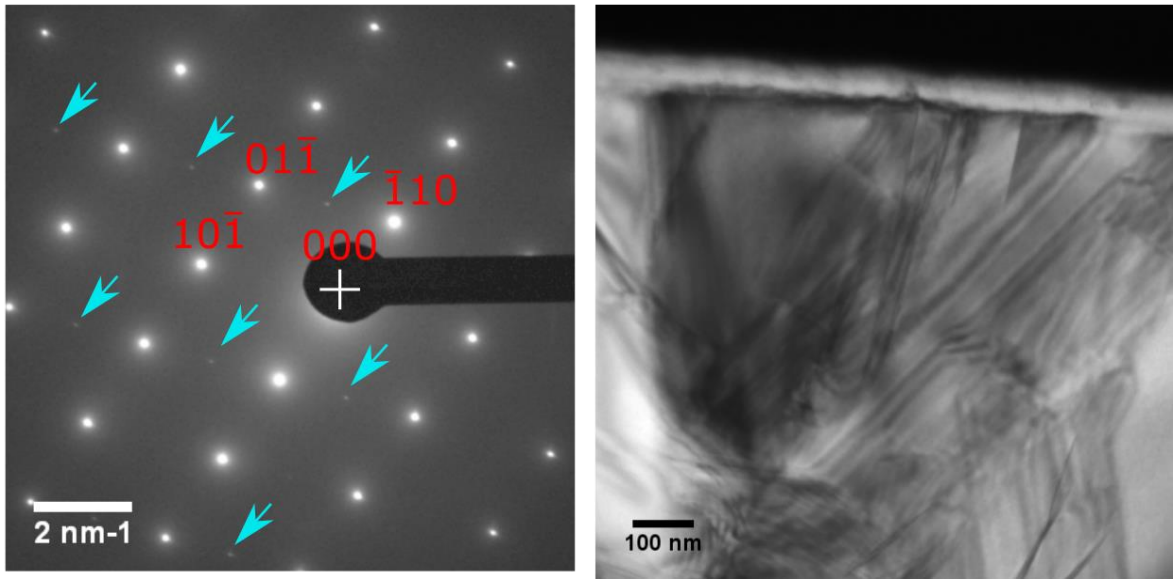

*Figure 14 SAED from a set of faults near the irradiated surface in the central heavily faulted grain of the 900°C irradiated TiBe$_{12}$ sample, which made superlattice spots as indexed in the text. The diffraction pattern has been enhanced by changing the gamma value to make the superlattice spots visible on this page, which are marked with blue arrows.*

#### 4.1.4.4 Relation between bubbles and faults

A low-magnification montage in ADF-STEM mode, showing some Z-contrast and some displacement contrast in which the faults are bright, is shown in Figure 15a. ADF-STEM images taken around the He concentration peak in the heavily faulted centre of the sample (Figure 15b and c) show small (~10nm) unfaceted bubbles with very bright contrast at some edges, surrounded by bright linear fault contrast. The very bright bubble edges do not correspond to concentrations of any high-Z element in EDS. This bright contrast is likely to be localised displacement, perhaps from structure rearrangement on the bubble walls. It may also be material that has been displaced as bubbles under pressure near the walls of the FIB resection have burst through to the new surface shortly after sample preparation.

A montage of ADF-STEM images was taken on-zone to encompass the whole sample, shown in Figure 15a. Using knowledge of the fault plane being {111}, the zone axis being [111], and the angles between the fault planes being 72° in the crystal structure, the sample thickness can be estimated at ~150 nm from projected widths of faults inclined to the sample plane, assuming faults run from top to bottom of the TEM sample and do not terminate part-way through the thickness. The area of the montage image corresponding to fault contrast in a band 600 nm tall containing the bubble-dense region around the He concentration peak can be segmented visually and calculated. The known fault geometries and sample thickness then allow the volume of the sample taken up by material in the immediate neighbourhood of faults to be estimated at 12% +/- 1.2% of the volume of the 600nm band. The number of bubbles in the image montage that appear in contact with fault contrast can then be counted as a proportion of the total number of bubbles; 90% of the bubble images overlap with fault contrast. If the bubbles were randomly placed with respect to the faults, one could expect that 12% of the bubble images would overlap with fault images; 90% overlap shows that the bubbles and faults are positively associated. There is some unknown error due to the assumption that faults are through-thickness, but the faults being part-thickness would make the "random bubble locations" estimate smaller than 12%, so the result stands. Work is ongoing to determine whether faults cause bubbles to precipitate, bubbles cause faults to develop, or some mutual root cause is behind both features.

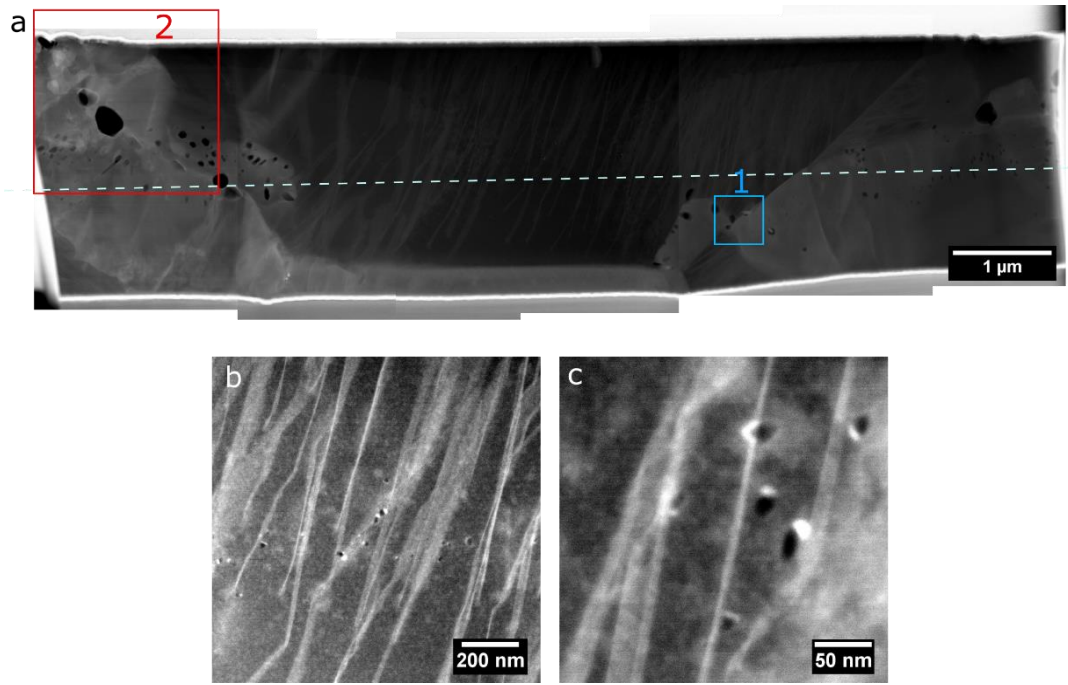

*Figure 15 (a) Low magnification ADF-STEM montage of TiBe$_{12}$ sample irradiated at 900°C. Bubbles appear dark, faults appear bright. The dotted line marks the position of the He zone. Boxes 1 and 2 are regions of interest in Figure 12. (b) ADF-STEM image from the 900°C irradiated TiBe$_{12}$ sample showing small bubbles and high fault density at the He concentration peak. The small bubbles are obscured by fault contrast in conventional CTEM. (c) higher magnification image of similar area.*

## 4.2 Cr beryllide samples

**4.2.1.** Sample RLC 1: CrBe$_{12}$ irradiated with 300keV He at 387°C

*4.2.1.1  Summary*

No bubbles are observed in this sample. Oxide particles of 100-200nm diameter are found with possible dislocation involvement indicated.

*4.2.1.2  Oxide particles*

A low magnification montage of this sample off zone is shown in Figure 16a. The unirradiated part of the specimen that He did not penetrate (towards the bottom of the image) is much more distorted than the irradiated part, shown by more bend contours. No faults are evident on this sample, but there is one particle around the He concentration peak depth. On zone, another feature can be seen on the irradiated side of the sample, shown in Figure 16c, a circular feature of diameter ~150nm. This is analysed further shortly. It can also be seen on zone that the unirradiated material under the He concentration peak (towards the top of the image) is speckled, as it was for the 387°C TiBe$_{12}$ sample, and the irradiated material is made of larger grains. The only defect found is an isolated dislocation structure downstream of the He band, towards the left of Figure 16, which does not appear to have bubbles attached on either side of focus.

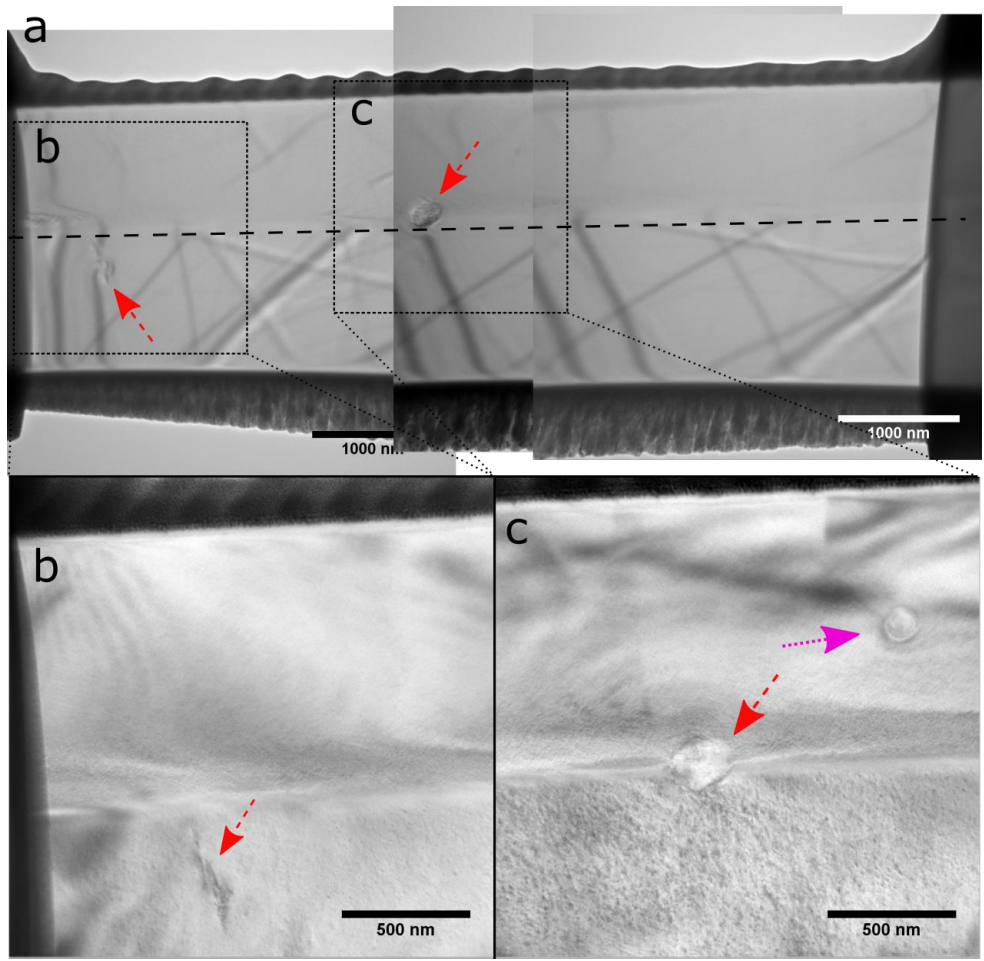

*Figure 16 a) Low magnification off-axis montage of CrBe$_{12}$ sample irradiated at 387°C. The He concentration peak is marked with a dashed line; the two features marked with red dotted arrows are a dislocation arrangement (left) and a particle (centre). Areas b and c are shown in on-zone images (b) and (c). In image (c) the particle (red/dashed arrow) is seen and another, circular object (pink/dotted arrow) which is visible only on zone.*

The particle at the He concentration peak depth is further analysed in Figure 17. In defocus series (a-c), long thin objects (approx. 3-4 nm wide and 30-60 nm long) are seen on the He concentration peak position, that have pale contrast at negative defocus, dark contrast at positive defocus (e.g. the feature at the end of the red arrows). These resemble long bubbles or perhaps platelets as seen in covalent compounds and some metals [18]. The long axes of their projected images mainly align closely with the $[\bar{3}\bar{2}1]^*$ reciprocal lattice direction on the diffraction pattern, which likely places their long axes or flat faces on some crystal plane of type $[\bar{n}\bar{m}1]$ where n and m are close but not necessarily equal non-zero integers. They are, however, not exactly parallel to each other. The +400nm overfocus images from the He concentration peak on the opposite side of the particle in Figure 17(d-f) show a band of smaller linear objects pointing in a different direction from the objects in (a-c). ADF STEM and EDS mapping shown in image set (g) shows the particle itself to be oxide, probably BeO given that it coincides with a hole in the Cr map. Image (d) also shows a rounded, slightly faceted object ~200 nm in diameter around the particle; this shows darker (less mass-thickness) in ADF STEM and partially brighter in the oxygen EDS map, indicating it contains oxide but is thinner than the main particle. It is not known at this time whether the thinner particle grew at the same time as the main particle or after irradiation and FIB sample preparation. The bright spot in the ADF STEM image is Mo, but this is probably sputtered from the support grid during FIB.

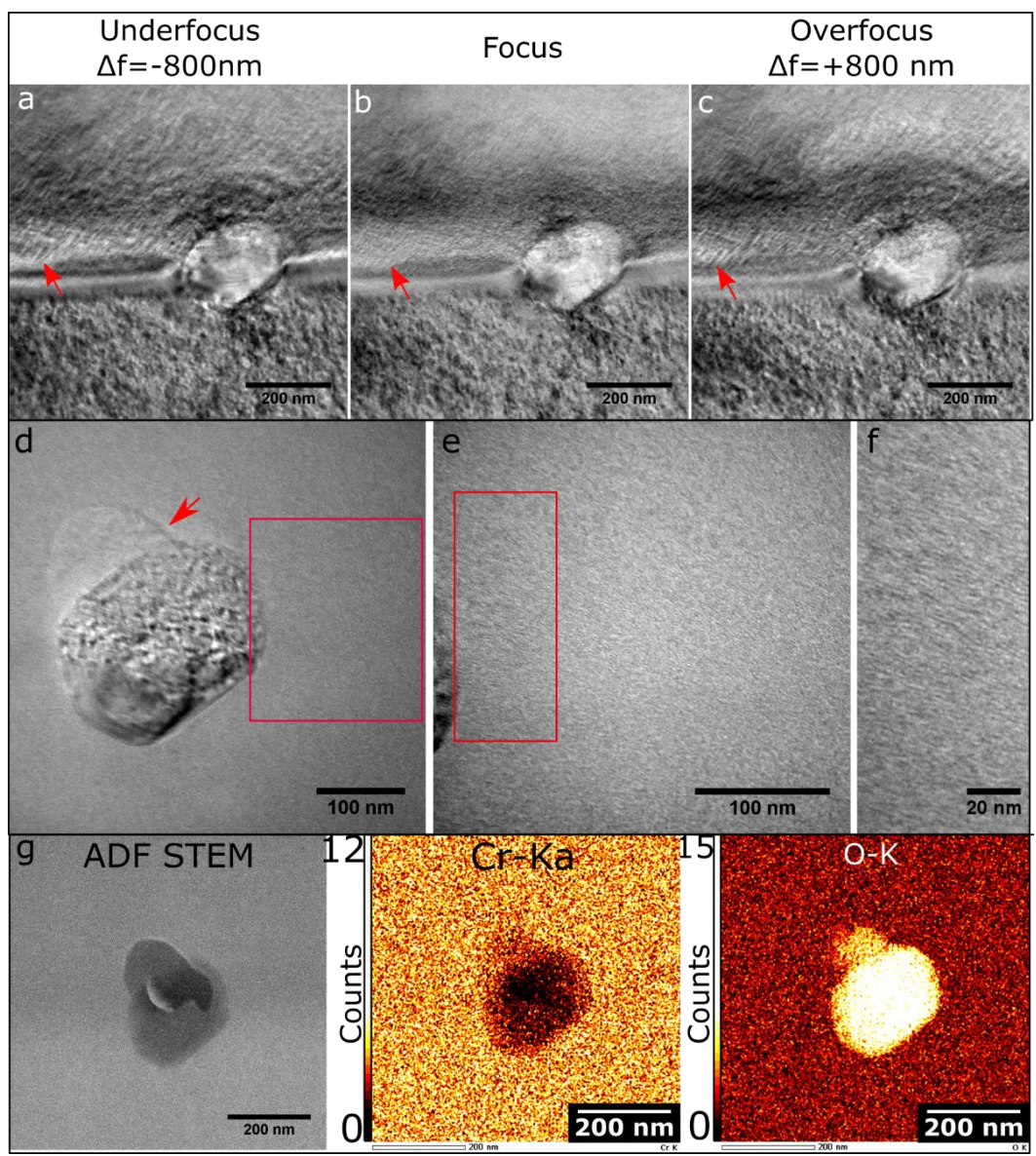

*Figure 17 Images of particle at He concentration peak of CrBe$_{12}$ sample irradiated at 387°C. (a-c) Focal set of on-axis bright field images. The red arrows are at equivalent points on each image showing an example of the flat features which invert contrast either side of focus, possibly flat bubbles or platelets. (d) A feature is seen attached to the particle that is more obvious at defocus but does not simply contrast reverse (marked by arrow). (e) the region in the box in d; (f) the strip marked in e, showing an increase in speckling across the He band on the scale of single nm. (g) ADF STEM image and EDS maps showing a faceted oxide particle and a thinner or less densely oxidised region associated with it.*

The most accessible zone axis was $[\bar{1}33]$. Bright field/dark field images of the particle and the circular object upstream of the He concentration peak (arrowed magenta/dotted in Figure 16c) BF/DF images of the circular object are shown in Figure 18a. The contrast here is typical of a dislocation loop, especially in the $[01\bar{1}]$ dark field image where the dark/light contrast switches as an extinction contour crosses the loop. Analysis by STEM EDS, however, shows that this object contains O, as shown in Figure 18b, and matches a gap in the Cr map (not shown). It also partly contains Mo; this is probably redeposition from the grid that has adhered preferentially to this object, implying that it is on the surface. Its visibility in CTEM only on zone, and bright contrast in the $g = \bar{3}0\bar{1}$ DF images suggests it has an orientation relationship with the matrix in which an oxide diffraction spot coincides with the $\bar{3}0\bar{1}$ spot of the CrBe$_{12}$ matrix. The presence of dark-light linear contrast around the circumference in other DF CTEM images may be due to a dislocation around the precipitate; this may be from deformation under stress from He implantation or thermal expansion mismatch, or a misfit dislocation if the interface is semi-coherent with misfit, or the precipitate acting as a point defect sink for damage at this depth further from the He concentration peak. Alternatively it may be a flat, circular bubble on the $(\bar{1}33)$ plane that was exposed to the air during sample preparation, gathered redeposited Mo and then oxide on exposure to air, but is no longer a bubble so does not show contrast inversion either side of focus off-zone. The structure of oxides in $CrBe_{12}$ is not yet known so finding the structure and the character and orientation of their interfaces with the matrix is required before more can be deduced about this object.

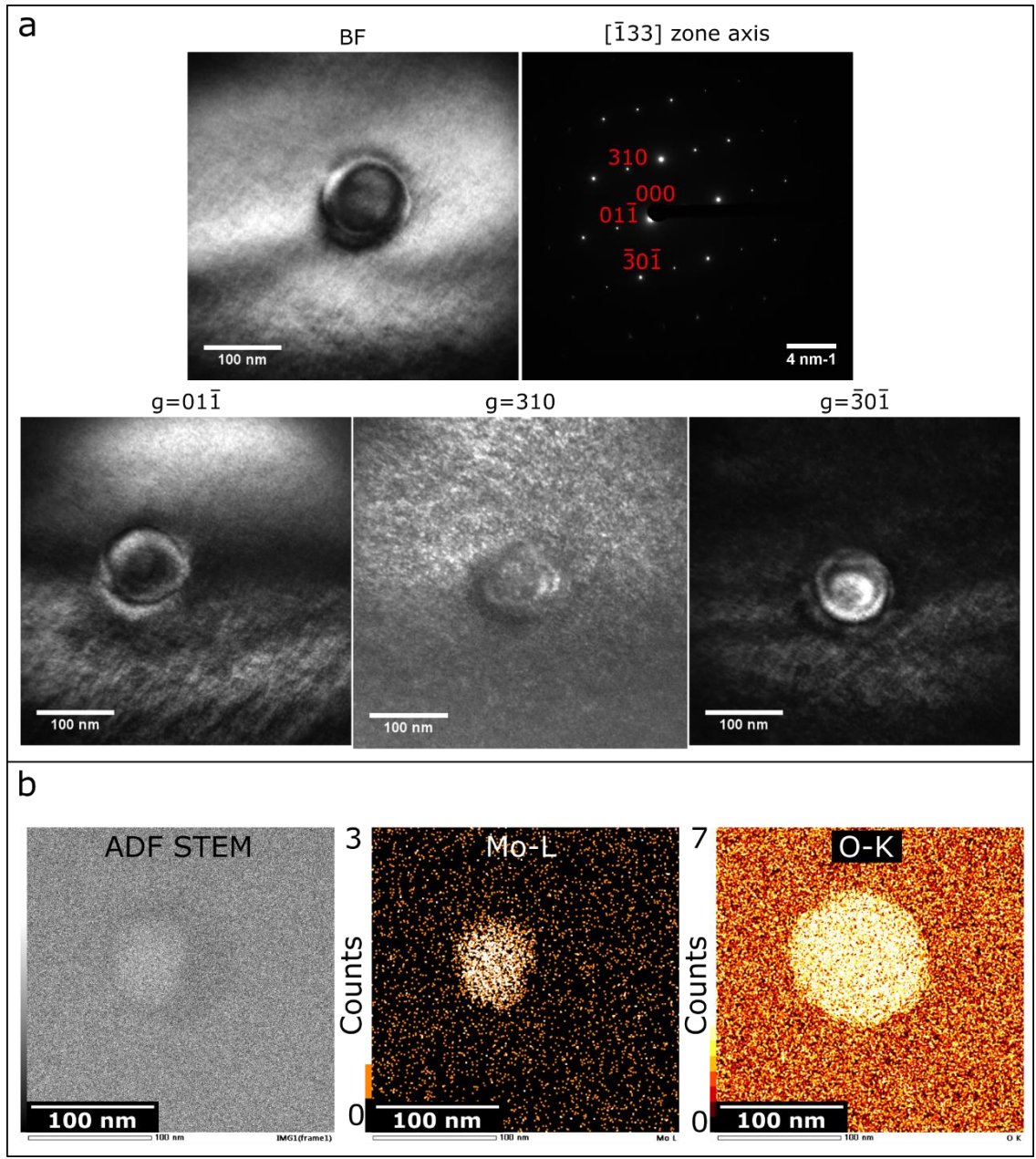

Figure 18 (a) BF/DF imaging of the circular object upstream of the He concentration peak in the 387°C irradiated $CrBe_{12}$ sample. (b) ADF STEM and EDS mapping from the object.

### 4.2.2. Sample RLC 2: CrBe$_{12}$ irradiated with 300keV He at 480°C

#### 4.2.2.1 Summary

Bubbles are not observed in this sample; speckling originating from Cu contamination is seen, as in some of the TiBe$_{12}$ samples. An arrangement of planar faults is found with not striped but swirled contrast, atypical of stacking faults in TEM.

#### 4.2.2.2 Absence of bubbles; Cu contamination

In this sample, more faults are observed which partially run along the He concentration peak, as shown in Figure 19, a montage of on-zone images. The dashed line marks the He concentration peak, the red arrows indicate faults that change direction or terminate at it. No bubbles are seen over a wide range of magnification and defocus conditions, but there is speckled contrast over the sample, identifiable in STEM-EDS as copper contamination. STEM-EDS did not reveal any precipitates in this particular FIB sample.

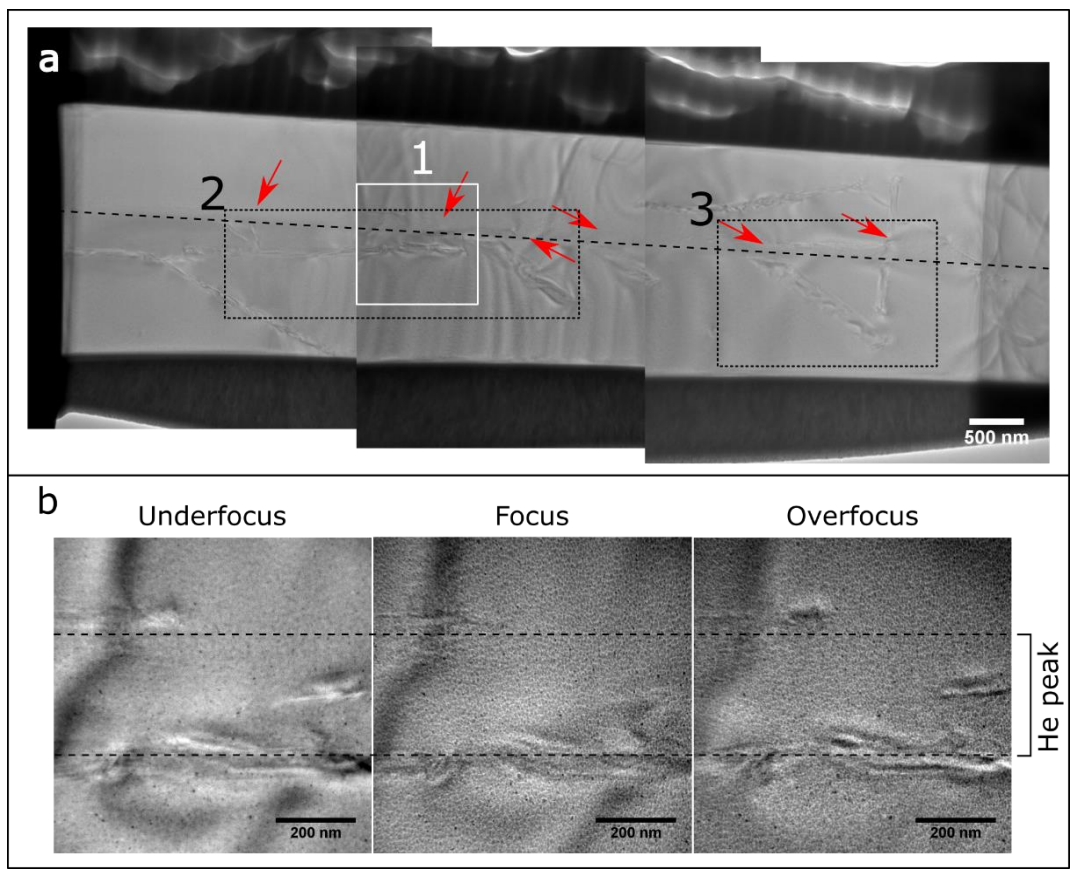

*Figure 19 (a) Low magnification montage of on-zone images of CrBe$_{12}$ sample irradiated at 480°C. The dotted line marks the He concentration peak, red arrows mark locations where faults interact with the strain around the He concentration peak. (b) Focal set of images at the He band from box 1 in (a). Underfocus/overfocus are -/+1200nm. The approximate location of the He band is labelled, judged from the contrast variation in (a). No evidence of bubbles was seen at these or other focus values. Images (b) are covered in fine dark speckled contrast, identified as Cu contamination using STEM EDS.*

#### 4.2.2.3 Planar faults

This sample is closest to the slightly esoteric zone axis $[\bar{1}37]$. Bright and dark field imaging of the faults is shown in Figure 20 for the fault arrangement in box 2 in Figure 19, and in Figure 21 for the number "4" shaped fault complex in box 3 in Figure 19. Both fault complexes are poorly visible in dark field images from g=121, to the extent that locating areas and assembling partial montages is impossible.

Both fault plane families from the TiBe$_{12}$ samples, {110} and {111}, would intersect at 79° and 101° when the structure is projected on the [13$\bar{7}$] zone axis, so would be indistinguishable from each other. The angles of intersection in the fault complexes in this sample, however, are close to 62° and 34° and do not match those angles, or projected angles that would be seen between any of the {110} and {111} family and Banerjee's observed fault plane of (001) [7]. This is a key difference between these faults and those found in TiBe$_{12}$. The dark field contrast on some of the faults seen in Figure 20 is also not striated but has a more swirled appearance; these may not be stacking faults at all but some other planar defect, perhaps microcracks.

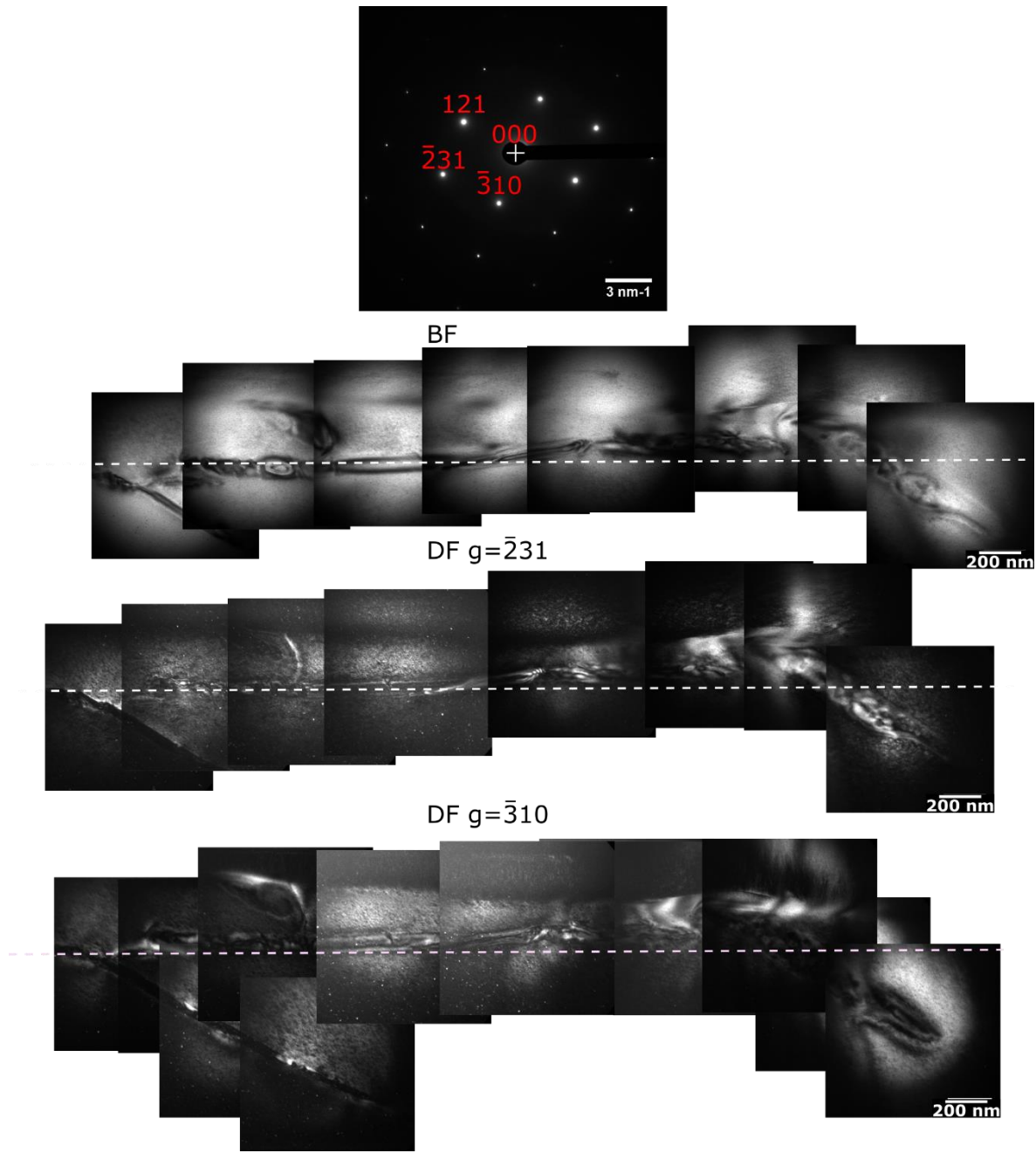

*Figure 20 BF/DF imaging from CrBe$_{12}$ sample irradiated at 480°C, showing long defect complex in box 2 in Figure 19. The structure is nearly invisible in the third DF spot g=121. White dashed lines show position of the He concentration peak.*

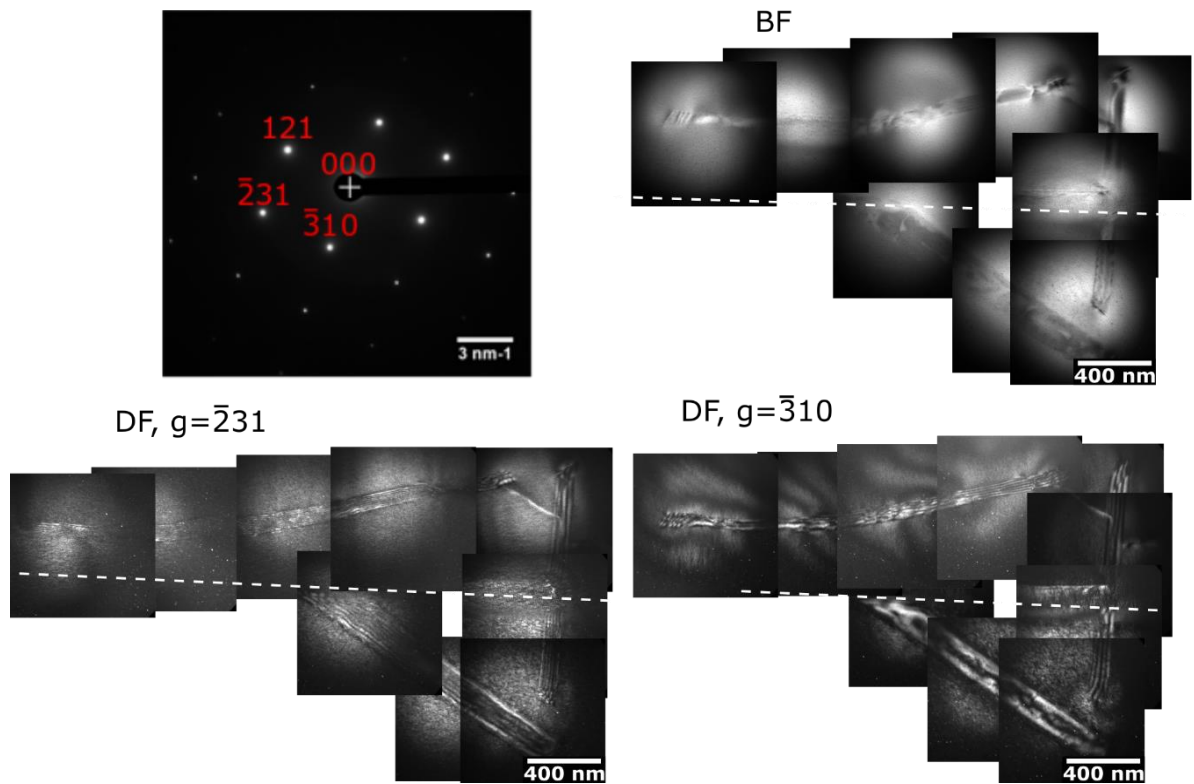

*Figure 21 BF/DF imaging from CrBe$_{12}$ sample irradiated at 480°C showing fault complex with numeral "4" shape as indicated in box 3 in Figure 19. The defects are invisible in g=121. White dashed lines show approximate position of He band.*

### 4.2.3. Sample RLC 3: CrBe$_{12}$ irradiated with 300keV He at 600°C and prepared by re-FIB

#### 4.2.3.1 Summary

Bubbles of 1nm size are found in the He concentration peak in this sample. Oxide particles are also found, but no faults in this particular FIB section.

#### 4.2.3.2 Bubbles

Defocused TEM images shown in Figure 22 reveal a bubble band of small bubbles on the scale of 1 nm above the top of the speckling, at the He concentration peak. The bubble band is of comparable width to that seen in the TiBe$_{12}$ sample irradiated at this temperature.

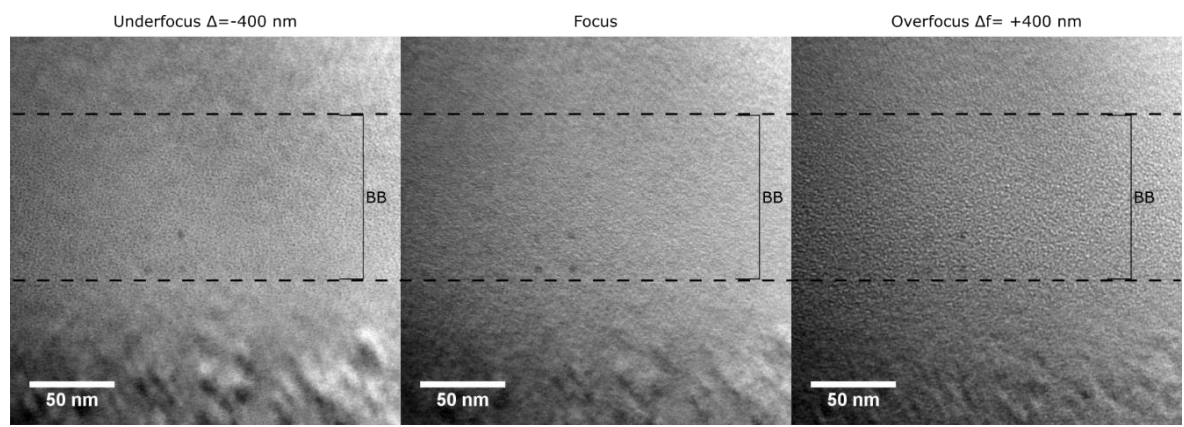

*Figure 22 Focal set of images showing small bubbles in CrBe$_{12}$ sample irradiated at 600°C. Bubble band is marked by BB.*

### 4.2.3.3 Oxide particles

This sample, similar to the 387°C irradiated TiBe$_{12}$ sample, appears speckled below the irradiated area but single crystal in the irradiated part as shown in Figure 23. Diffraction patterns taken in the smooth and speckled regions are identical, both showing one strong pattern from the CrBe$_{12}$ grain and diffuse contrast from surface oxide and/or redeposition from the Cu FIB support grid. Be-rich oxide precipitates are seen upstream and downstream of the He concentration peak.

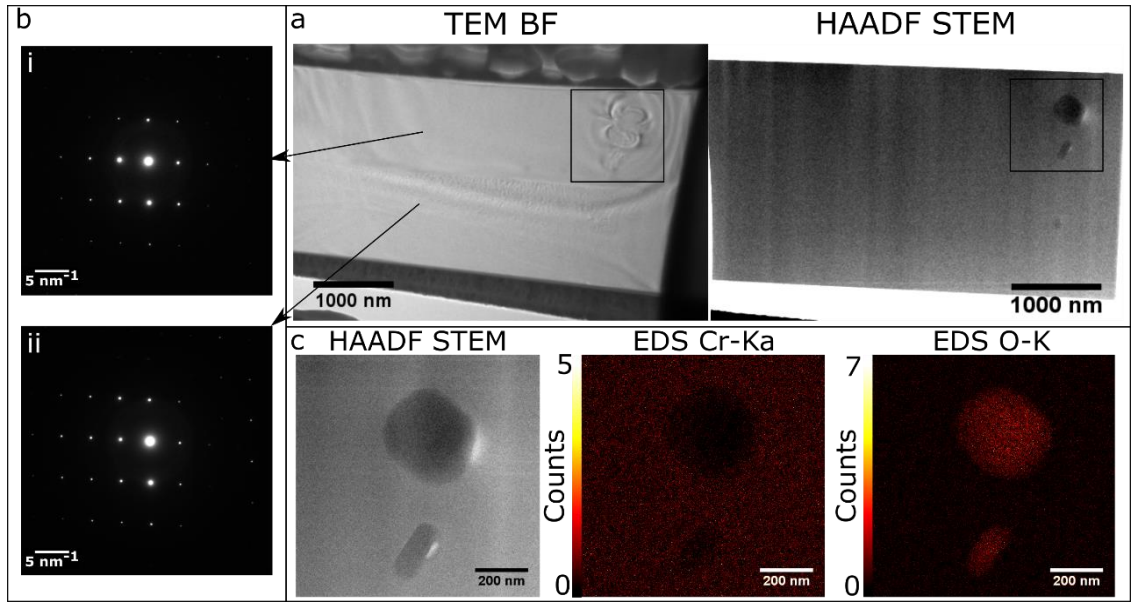

Figure 23 (a) Low magnification TEM BF and HAADF-STEM images of CrBe$_{12}$ irradiated at 600°C. (b) Diffraction patterns from (i) the smooth area above the bubble band, (ii)the speckled area below, both showing one main grain and diffuse contrast from surface oxide or FIB-induced surface amorphization. (c) Higher magnification HAADF-STEM and EDS maps showing oxide precipitates.

### 4.2.4. Sample RLC 4: CrBe$_{12}$ irradiated with 300keV He at 900°C

#### 4.2.4.1 Summary

As for TiBe$_{12}$, this sample is very different to those irradiated at lower temperatures. Large bubbles (10-100 nm) are found, including some nucleated on oxide precipitates. Some bubble surfaces are decorated with Cr, some of which is oxidised.

#### 4.2.4.2 Bubbles

This sample was prepared with a smaller electron-transparent area because previous samples with larger thin areas bent on thinning, possibly due to higher internal stress in the slab irradiated at 900°C. Low magnification images of the thin area are shown in Figure 24a. There is a clear band of bubbles across the He concentration peak depth, unlike from the lower irradiation temperatures. Larger bubbles (up to 100 nm) are found at the He concentration peak and smaller bubbles (~20 nm) further from it. Interestingly, smaller bubbles are found not only upstream of the peak He depth but also downstream. This implies either some spread of He destination due to diffusion at this high temperature and possibly substantial vacancy diffusion to form voids; voids nucleate on He bubbles in many materials e.g. [19]. Measuring the concentration of He in bubbles using high energy resolution EELS [20], and looking for depth correlation within the cross section, would provide interesting data in future. Both sizes of bubbles are faceted rather than spherical. A higher magnification TEM focus set in Figure 24b shows a precipitate in the bubble band of diameter ~50 nm and on which a bubble has nucleated and grown. The same precipitate is analysed in EDS in

Figure 24c, which shows it is a Cr-poor (and therefore Be-rich) oxide. In the same figure, we see some of the bubble margins are enriched in Cr and some of them slightly in O, but not every hotspot on the Cr map has a counterpart in the O map. This implies there is Cr segregation to the bubble edges during irradiation which subsequently oxidises after FIB sample preparation when they are exposed to air. This is a significant difference from the TiBe$_{12}$ specimens in which there is no elemental enrichment around the bubbles.

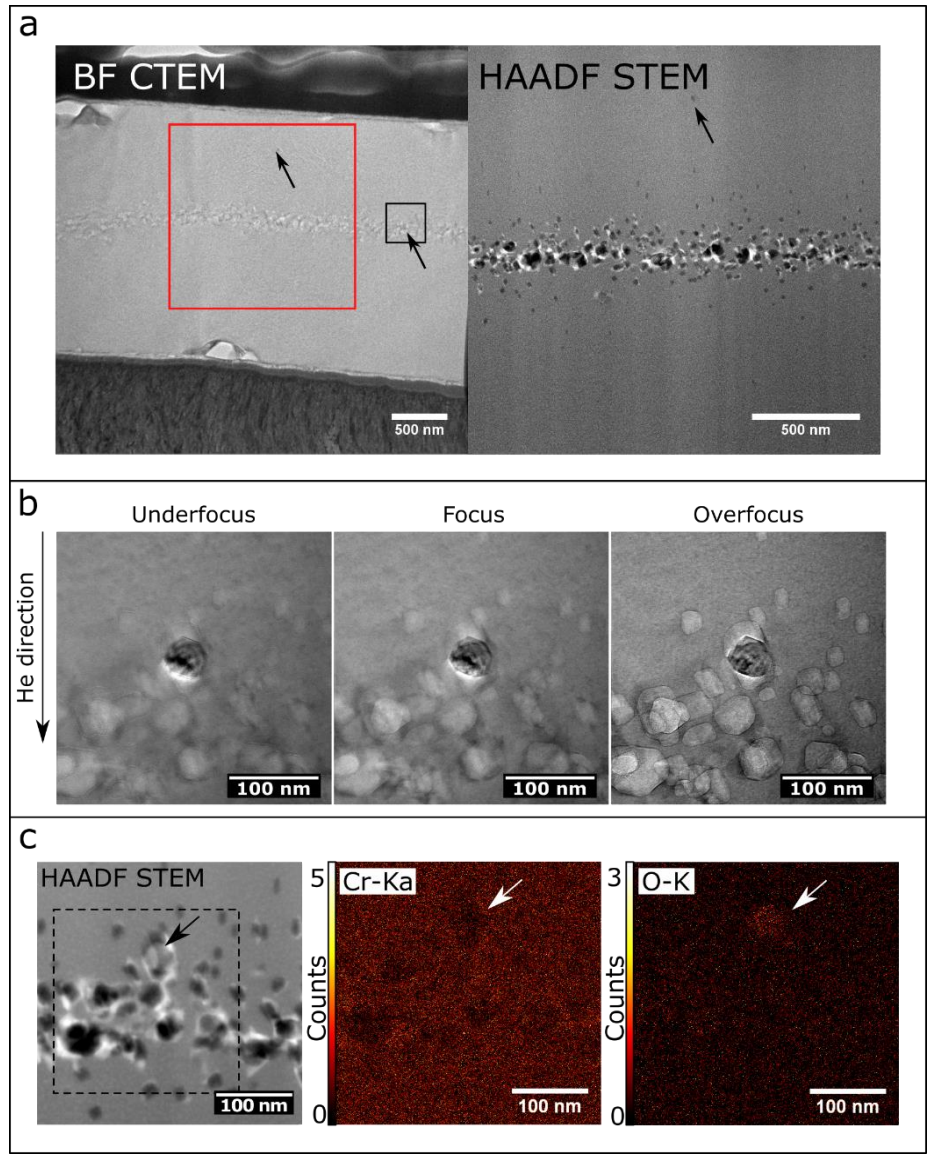

*Figure 24 (a) Low magnification images of 900°C irradiated CrBe$_{12}$ sample. A band of bubbles is clearly visible around the predicted He peak depth. The arrows show precipitates in the bubble belt and upstream of it. (b) Higher magnification off-axis TEM focal set of the precipitate in the bubble band; the ~50 nm precipitate is in a bubble that nucleated and grew around it. Underfocus/overfocus are -/+400nm. (c) EDS mapping of the precipitate in (b) which is Be-rich oxide; there is also Cr decorating the bubble edges.*

Imaging of the bubble band and precipitate in bright and dark field yields some interesting observations (Figure 25). Around some of the bubbles, fringes are found in the dark field image from g=231. The fringes in DF-CTEM coincide with the Cr decoration in EDS. This may imply some kind of orientation relationship between the Cr or Cr oxide and the CrBe$_{12}$ and a (semi-)coherent interface is implied, containing defects which cause the fringes; further work is required to find out.

In this sample no grain boundaries were captured, so it is not possible to compare the balance of bubble locations to that in TiBe$_{12}$ irradiated at the same temperature at this stage.

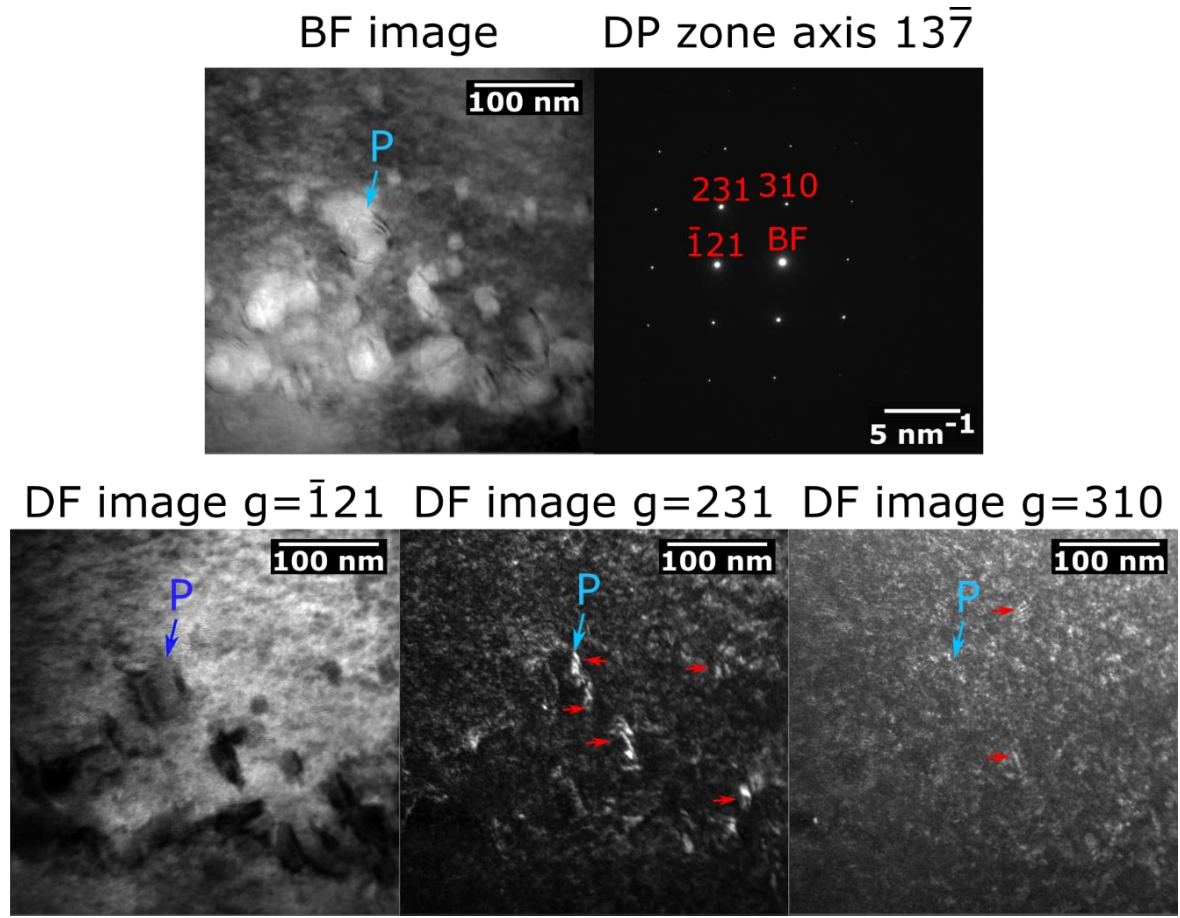

*Figure 25 BF/DF images from CrBe$_{12}$ sample irradiated at 900°C of bubble band with precipitate. Precipitate is indicated by blue arrow and letter P. Red arrows on g=231 and 310 dark field images show fringes around bubbles.*

## 5. Discussion

The origins of the planar defects and dislocations found in this study cannot be definitively attributed to the helium implantation. They may simply be the result of deformation, whether from manufacture, sample preparation, thermal expansion/contraction stress, or from the introduction of a large concentration of interstitial atoms into the structure. These faults are, however, a newly found feature of the beryllides under study; the {110} and {111} habit planes in TiBe$_{12}$ dependent on temperature, and the apparent absence of a habit plane and swirled appearance of the faults in CrBe$_{12}$ have not been reported in the literature before. Their relationship with irradiation will only be fully clarified by future experiments.

### 5.1 Summary table and trends with temperature

The results are summarised in Table 1 for quick reference while reading the discussion.

*Table 1 Summary of results. "None seen" acknowledges none of the given feature were seen in the sample studied but this does not rule out their existence due to small volumes analysed so far in comparison with the sparseness of the feature in question.*

| Material | Temp. (°C) | Bubbles | Defects | Precipitates |
| --- | --- | --- | --- | --- |
| TiBe$_{12}$ | 387 | None | Isolated planar faults on {110} planes; dislocations $b = \frac{1}{2}[111]$ | None seen |
| | 480 | None | Group of parallel planar faults on {110} planes that locally show [001] ZA pattern | None seen |
| | 600 | Small (~1nm) | Complex of planar faults on different {110} planes | None seen |
| | 900 | Faceted, 50-100nm in grain interiors; rounder, >200nm on grain boundaries; no elemental segregation | High density of intersecting planar faults on {111} planes, associated with grain interior bubbles | Oxide at grain boundary triple point |
| CrBe$_{12}$ | 387 | None | None seen | Oxides |
| | 480 | None | Complex of intersecting planar faults on no particular plane group | None seen |
| | 600 | Small (~1nm) | None seen | Oxides |
| | 900 | Semi-faceted, 20-100nm in grain interiors, gb not seen; Cr segregation with some oxide. | None seen | None seen |

Both TiBe$_{12}$ and CrBe$_{12}$ developed bubbles at 600°C of ~1nm diameter in a band of ~100nm width at the He concentration peak. Both beryllides also developed larger bubbles at 900°C, which nucleated on oxide precipitates. Grain boundaries were also nucleation sites for the largest bubbles in TiBe$_{12}$, and for some bubbles or voids away from the He concentration peak, but no grain boundaries were captured in the CrBe$_{12}$ 900°C sample for comparison. In CrBe$_{12}$, some bubbles had decoration of Cr or a more Cr-rich phase such as CrBe$_2$ [21] on the interior surfaces; some of this Cr-rich material was oxidised, possibly after FIB sample preparation. This is similar to the role of Cr in preventing corrosion in stainless steels, and may give CrBe$_{12}$ improved corrosion resistance over TiBe$_{12}$.

### 5.2 Dislocations in TiBe$_{12}$

Dislocations were seen in two of the TiBe$_{12}$ samples: the long curved dislocations next to the fault in the 387°C sample, and at the ends of the faults in the 480°C sample. In Banerjee et. al.'s previous work on defects in beryllides, Burgers vectors of $\frac{1}{2}\langle 111 \rangle$ were derived from **g.b** analysis.

In this study's TiBe$_{12}$ sample irradiated at 387°C, dislocations were strongly visible in $g = 1\bar{1}0$ and $g = 200$ but showed low contrast in $g = 110$ (strict invisibility at $\boldsymbol{g}.\boldsymbol{b} = 0$ is only the case for materials with cubic symmetry, but in general contrast is dependent on the magnitude of $\boldsymbol{g}.\boldsymbol{b}$ [22]). This is consistent with a Burgers vector of $\boldsymbol{b} = \frac{1}{2}[1\bar{1}1]$. In the 480°C sample, dislocations were strongly visible in $g = 12\bar{1}$ and $g = 200$ but showed low contrast in $g = 10\bar{1}$; this is also consistent with a Burgers vector of $\boldsymbol{b} = \frac{1}{2}[111]$. In respect of dislocations, this work's findings are similar to those of Banerjee et. al. [7]

### 5.3 Planar faults in TiBe$_{12}$

The previous work by Banerjee et. al. [7] found antiphase boundaries with displacement vector $\boldsymbol{R} = \frac{1}{2}\langle 110 \rangle$. Those antiphase boundaries were an artefact of arc melting producing metastable Ti$_2$Be$_{17}$ as a side product; these samples were prepared by HIP of extruded powder with longer at high temperature and a slower cooling rate, which compared with arc melting was found to produce fewer side phases [9]. Those antiphase boundaries were curved; the faults found in the current work are straight. Those researchers also found the superlattice-creating planar defects on (001) planes with displacement vector $\boldsymbol{R} = \frac{1}{2}\langle 011 \rangle$ which made a single plane of Ti$_2$Be$_{17}$ unit cells. The faults found in this study are on a plane inconsistent with that type of planar defect. Nevertheless, the displacement vectors in Banerjee et. al.'s stacking faults are good candidate displacement vectors for the faults found here, as they represent energetically favourable atomic configurations on the fault plane.

Similar to the **g.b** criterion for dislocations, there is a visible/invisible (or for non-cubic materials, strongly/weakly visible) criterion for planar faults. The displacement vector **R** is involved in the phase factor $\alpha = 2\pi \boldsymbol{g} \cdot \boldsymbol{R}$ where **g** is the diffraction vector of the dark field image. The contrast of the fault depends on the phase shift α that it imparts on the electron wavefunction passing through, which cycles around a sine function; it has maximum contrast when $\alpha = \pi$ or an odd integer multiple of π, the maximum of the sine function where the scattered wave has maximum phase shift from the unscattered wave, and zero or minimum contrast when $\alpha = 2\pi$ or an even integer multiple of π, i.e. the phase shift has brought the scattered wavefunction back in phase with the unscattered wave [22]. We therefore calculate $\boldsymbol{g} \cdot \boldsymbol{R}$ for each diffraction spot **g** that we have taken dark field images from, multiply by 2π, and construct a table of α values for each possible fault displacement vector **R**, shown in Table 2.

|  | g | **R** displacement vectors for APB type fault | | | **R** displacement vectors for plane of Ti$_2$Be$_{17}$ type fault | | |
|---|---|---|---|---|---|---|---|
|  |  | $\frac{1}{2}[110]$ | $\frac{1}{2}[\bar{1}10]$ | $\frac{1}{2}[1\bar{1}0]$ | $\frac{1}{2}[011]$ | $\frac{1}{2}[0\bar{1}1]$ | $\frac{1}{2}[01\bar{1}]$ |
| 387°C RLB1 | $1\bar{1}0$ | 0 (x) | -2π (x) | 2π (x) | -π (v) | π (v) | -π (v) |
|  | 110 | 2π (x) | 0 (x) | 0 (x) | π (v) | -π (v) | π (v) |
|  | 200 | 2π (x) | -2π (x) | 2π (x) | 0 (x) | 0 (x) | 0 (x) |
| 480°C RLB2 | $0\bar{2}0$ | 2π (x) | 2π (x) | -2π (x) | 2π (x) | -2π (x) | 2π (x) |
|  | $12\bar{1}$ | 3π (v) | π (v) | -π (v) | π (v) | -3π (v) | 3π (v) |
|  | $10\bar{1}$ | π (v) | -π (v) | π (v) | -π (v) | -π (v) | π (v) |
| 600°C RLB3 | $12\bar{1}$ | 3π (v) | π (v) | -π (v) | π (v) | -3π (v) | 3π (v) |
|  | $2\bar{1}\bar{1}$ | π (v) | -3π (v) | 3π (v) | -2π (x) | 0 (x) | 0 (x) |
|  | $1\bar{3}0$ | -2π (x) | -4π (x) | 4π (x) | -π (v) | π (v) | -π (v) |

Table 2 Visibility pattern table for dark field reflections used to image planar faults in TiBe$_{12}$ samples RLB1-3. Central cells contain phase factor α and letter in brackets indicates the predicted contrast of the given fault at this **g**, V to indicate visible or X to indicate invisible or weakly visible.

We then compare these predictions with observations on the faults seen.

- In the 387°C sample RLB1 only one fault is imaged, running top to bottom in Figure 6 and in the precession map of Figure 5. This is weak in the $\boldsymbol{g} = 1\bar{1}0$ dark field image, strong in the 110 image, weak in the 200 image, or in the notation of Table 2, x v x; none of the possible displacement vectors in the 387°C RLB1 row match this, so according to the diffraction data, this fault does not fit either of the two fault types found by Banerjee et. al. [7].
- In the 480°C sample RLB2 a set of parallel faults was seen emerging from the corner grain in Figure 8; these are weak in the $\boldsymbol{g} = 020$ dark field image, strong in $\boldsymbol{g} = 12\bar{1}$ and strong in the $\boldsymbol{g} = 10\bar{1}$ image, pattern x v v. This fits all of the possible displacement vectors, so these faults could be either of the fault types found in [7] or some different type.
- In the 600°C sample RLB3 the faults in the triangular arrangement of Figure 10 have different patterns of visibility in the dark field images, so must be treated as separate entities. Fault A, for example, is strong in the $\boldsymbol{g} = 12\bar{1}$ dark field image, weak in $\boldsymbol{g} = 2\bar{1}\bar{1}$ and strong in $\boldsymbol{g} = 1\bar{3}0$, so has a visibility pattern v x v. Visibility patterns for the whole set are: A vxv, B xvv, C vvv, D vvv, E xvv. Fault A matches any of the $\boldsymbol{R} = \frac{1}{2}\langle 011 \rangle$ family of faults that create a unit cell of Ti$_2$Be$_{17}$; the remaining faults fall into two similar pairs, B/E and C/D; both pairs match neither fault family.

From this analysis, it can be inferred that there are other possible planar fault types in TiBe$_{12}$ in addition to those seen in [7]. They may be different because this sample has been irradiated; conversely, because TEM studies necessarily only observe a small volume of material, the two studies may have sampled partly-overlapping small subsets of a mutually applicable defect landscape in TiBe$_{12}$. Areas for further work include finding models for additional fault types that would explain the remaining visibility patterns here, and simply doing more DF imaging on this material to gain a larger population of data.

This **g**.R analysis was not undertaken for the {111} plane faults in 900°C sample RLB4, because those faults were too dense and overlapping to state whether individual faults are strong or weak in a given image; that too requires more images and further work. The faults in CrBe$_{12}$ are not fringed, so it is unlikely they follow the same contrast equations as those that describe dark field images of conventional planar faults.

It is also productive to consider the broader origins of these faults and their change with irradiation temperature. There are three influences on the material in these experiments: the stress due to implanted interstitial He; the concentration of point defects from irradiation, particularly vacancies; and the development of bubbles. It is possible that the faults found are independent of the irradiation and are merely a response of the beryllide materials to stress – from manufacture, thermal expansion and contraction and/or from implanted He. It is also possible that they were not present in the bulk material and developed during stress relaxation when the FIB sections were prepared. Even if the faults are not a result of the irradiation, they are a characteristic of beryllides and are relevant to understanding their mechanical properties and irradiation behaviour. Faults of this nature have not been seen previously in non-irradiation-related characterisations of the beryllides [7], however.

In terms of temperature, bubbles are normally the lowest energy state of gas atoms such as He in a solid material e.g. [2], [23], [24], but there are several kinetic activation barriers to their development. There has to be a nucleation point, for example a grain boundary, precipitate, defect or cluster of irradiation-induced vacancies. Once nucleated and grown past the critical survival size [25], more He has to diffuse to the growing bubble, accompanied by movement of further vacancies if He-vacancy clusters such as those predicted in [6] are involved; a bubble is stable, however, it is a sink that absorbs vacancies [19]. During growth the surrounding material has to deform to accommodate the change in physical size and shape of the bubble. Temperature affects the ability of He to diffuse, the mobility of vacancies, and in some materials the possible deformation mechanisms.

The change of defect structure in TiBe$_{12}$ from isolated {110} planar defects to interacting {111} planar defects may therefore have two origins, or a mixture of the two: the increase in temperature supplies the larger activation energy for creating of a {111} fault type that is more efficient at relieving stress; and/or, a higher temperature enables He and vacancies to diffuse faster to the bubbles during the irradiation which then gives a higher bubble growth rate that is enough to overcome an energy threshold to trigger cascades of {111}-type faults. These are theoretical starting points for further investigation. Understanding the nature of the different fault types is key in future work; their displacement vectors, the arrangements of atoms on the fault planes, their activation energies for creation, how they move (e.g. whether dislocations are involved) and therefore which parts of an applied stress tensor they can relieve, the activation energies for the different types of motion in and out of their fault plane, and whether the faults themselves can act as nucleation sites or diffusion paths for He.

This is an area in which further theoretical work would be very advantageous, as well as more experimental data.

### 5.4 Bubble faceting

The faceted bubbles in the TiBe$_{12}$ sample irradiated at 900°C (Figure 11b) have plane angles to the horizontal image axis including 9°±1°, 69°±3° and 127°±3°, from a population of 16 bubbles in the focal set images. When a model in CrystalMaker is oriented to match the rotation-corrected diffraction pattern, there are no low-index planes that fall in these ranges; in particular the {110} plane family and {111} plane family do not make these angles in projection. Tracing out these angles on a crystal structure projection does not produce obvious lines from one atom to another to suggest any simple physical relationship between the bubble facets and the crystal structure. It is difficult to assess planes from a 2D projection, however, and future work should involve tomography (X-ray or electron depending on scale) to reconstruct 3D shapes of bubbles and measure their facet planes directly.

Bubbles in the 900°C irradiated CrBe$_{12}$ sample are projected into the images with rounder edges and corners than in TiBe$_{12}$, and measuring facet angles is not possible with reasonably small error. The grain zone axis in this sample of $[13\bar{7}]$ does not lend itself to low-index crystal facets being end-on in projection giving this rounded facet projection when imaged on-zone. In general, analysis of bubble facet planes in both these materials requires a somewhat more focused further study.

### 5.5 Points for further study

Throughout this paper, features have been analysed and approaches suggested for further analysis. These are collated here for easier reference.

- Characterising the shapes and facet planes of faceted bubbles in 3D through X-ray and/or electron tomography
- Analysis of a larger population of faults at different zone axes in more samples to gain data about their displacement vectors
- Finding accurate temperature and nature of the transition between fault types in TiBe$_{12}$, understanding the atomic configurations on the fault planes and their interactions with each

- other and with bubbles, point defects (vacancies and self and gas interstitials) and dislocations, including modelling studies
- Determining the structure of the faults in CrBe$_{12}$ that cause the swirled contrast observed
- Determining oxide structures and interfaces in CrBe$_{12}$
- Control studies in which samples are subjected to heat treatment, stress without irradiation, and equivalent sample preparation through re-FIB to disambiguate which microstructural effects occur only with He irradiation
- Irradiating these materials to produce point defect damage but not He implantation – either some kind of compositionally equivalent "self-ion" beam, or even a sufficiently high-current electron beam, given the low atomic mass of Be

## 6. Conclusions

The irradiation microstructures of titanium and chromium beryllides both feature the onset of fine bubble development near an irradiation temperature of 600°C and larger bubbles, some faceted, by 900°C. Both beryllides developed planar faults, though it is not known yet whether these were caused directly by irradiation or as an ordinary microstructural response to stress, including that from the high number of interstitial He and/or bubbles at the peak He depth and the effects of stress release upon sample thinning.

The planar faults that developed were different for the two beryllides. TiBe$_{12}$ developed planar faults, on {110} planes at 600°C and below, whereas fault plane changed to {111} by 900°C, and these faults were preferentially associated with bubbles. The displacement vectors partially corresponded to those found in previous studies of $\boldsymbol{R} = \frac{1}{2}\langle 011 \rangle$ and $\boldsymbol{R} = \frac{1}{2}\langle 110 \rangle$, but some faults were found whose **g.R** behaviour matched none of those displacement vectors. At the lowest temperature of 387°C, dislocations were seen that matched previously observed Burgers vectors of $\boldsymbol{b} = \frac{1}{2}\langle 111 \rangle$. The planar faults in CrBe$_{12}$ had marbled contrast on the fault faces, rather than the striped contrast normally seen in planar crystal defects such as stacking faults; their inter-fault angles were inconsistent and did not correspond to any particular family of crystal planes. The difference in fault types may be the beginning of an explanation for the difference in mechanical properties between the two beryllides [26], [21].

CrBe$_{12}$ high temperature bubbles were partially lined with Cr or a Cr-rich phase, some of which was oxidised. This was not the case for TiBe$_{12}$. CrBe$_{12}$ also contained oxide particles at all temperatures on which bubbles nucleated; this was only the case for TiBe$_{12}$ at 900°C. This conclusion should particularly come with attached caution because the small volumes of material involved.

This study has gone some way to fill the knowledge gap on beryllide microstructures by providing an initial characterisation of the defects and bubble types to be expected when irradiation damage and He are introduced into TiBe$_{12}$ and CrBe$_{12}$ in the tokamak breeder relevant temperature range. It now serves as a springboard for further studies into the origins (irradiation and/or strain), nature and behaviour of these microstructural features.

## 7. Acknowledgements

This project was funded by the Engineering and Physical Sciences Research Council, United Kingdom (EPSRC) under grant number EP/T027193/1. The construction of MIAMI–2 was funded by EPSRC grant number EP/M028283/1. Sample preparation was carried out at UKAEA's Materials Research Facility, which has been funded by and is part of the UK's National Nuclear User Facility and Henry

Royce Institute for Advanced Materials and the RCUK Energy Programme (grant number EP/T012250/1).

This work has been carried out within the framework of the EUROfusion Consortium, funded by the European Union via the Euratom Research and Training Programme (Grant Agreement No 101052200 — EUROfusion). Views and opinions expressed are however those of the author(s) only and do not necessarily reflect those of the European Union or the European Commission. Neither the European Union nor the European Commission can be held responsible for them.